\documentclass[]{spie}  

\usepackage{amsmath,amsfonts,amssymb}
\usepackage{graphicx}
\def\gsim{\;\lower.6ex\hbox{$\sim$}\kern-6.7pt\raise.4ex\hbox{$>$}\;}
\def\lsim{\;\lower.6ex\hbox{$\sim$}\kern-6.7pt\raise.4ex\hbox{$<$}\;}

\title{Stellar photometry with Multi Conjugate Adaptive Optics}

\author[a]{Giuliana Fiorentino}
\author[a,b]{Davide Massari}
\author[c]{Alan McConnachie}
\author[c]{Peter B. Stetson}
\author[d,e]{Giuseppe Bono}
\author[c,f]{Paolo Turri}
\author[c]{David Andersen}
\author[c]{Jean-Pierre Veran}
\author[a]{Emiliano Diolaiti}
\author[a]{Laura Schreiber}
\author[a]{Paolo Ciliegi}
\author[a]{Michele Bellazzini}
\author[b]{Eline Tolstoy}
\author[g]{Matteo Monelli}
\author[e]{Giacinto Iannicola}
\author[e]{Ivan Ferraro}
\author[e]{Vincenzo Testa}

\affil[a]{INAF-Osservatorio Astronomico di Bologna, via Ranzani 1, 40127, Bologna, Italy}
\affil[b]{University of Groningen, Kapteyn Astronomical Institute, NL-9747 AD Groningen, The Netherlands}
\affil[c]{Herzberg Astronomy and Astrophysics, National Research Council Canada, 5071 West Saanich Road, Victoria, BC V9E 2E7, Canada}
\affil[d]{Dipartimento di Fisica, Universit\'{a} di Roma Tor Vergata, Via della Ricerca Scientifica 1, 00133 Roma, Italy}
\affil[e]{INAF-Osservatorio Astronomico di Roma, Via Frascati 33, 00040 Monte Porzio Catone, Italy}
\affil[f]{Department of Physics and Astronomy, University of Victoria, 3800 Finnerty Road, Victoria, BC V8P 5C2, Canada}
\affil[g]{Instituto de Astrof\'{i}sica de Canarias, Calle Via Lactea s/n, E38205 La Laguna, Tenerife, Spain.}
\authorinfo{Further author information: (Send correspondence to G.F.)\\: E-mail: giuliana.fiorentino@oabo.inaf.it, Telephone: +39 051 2095318}

\begin{document} 
\maketitle

\begin{abstract}
We overview the current status of photometric analyses
of images collected with Multi Conjugate Adaptive Optics (MCAO) 
at 8-10m class telescopes that operated, or are operating, on
sky. Particular attention will be payed to resolved stellar population
studies. Stars in crowded stellar systems, such as globular
clusters or in nearby galaxies, are ideal test--particles to
test AO performance. We will focus the discussion on  
photometric precision and accuracy reached nowadays. We briefly 
describe our project on stellar photometry and astrometry of
Galactic globular clusters using images taken with GeMS at the Gemini
South telescope. We also present the photometry performed with DAOPHOT
suite of programs into the crowded regions of these globulars reaching
very faint limiting magnitudes K$_s\sim$21.5 mag on moderately large 
fields of view ($\sim$1.5 arcmin squared). We highlight the need for 
new algorithms to improve the modeling of the complex variation of 
the Point Spread Function (PSF) across the field of view. Finally, 
we outline the role that large samples of stellar standards plays 
in providing a detailed description of the MCAO performance and in 
precise and accurate colour--magnitude diagrams.
\end{abstract}

\keywords{Photometry, Adaptive Optics, Resolved Stellar Populations, Globular Clusters}

\section{INTRODUCTION}
\label{sec:intro}  

Resolved stellar populations are crucial diagnostics to improve our
understanding of the early formation and evolution of giant and dwarf
galaxies in the Local Group. The Local Group is an unique laboratory
to constrain cosmological simulations. The current cold dark matter
model predicts that large galaxies like Andromeda and the Milky Way
assembled at early epochs from merging of several dwarf
galaxies. Solid constraints on the early formation and evolution of
giant galaxies rely on deep and accurate color magnitude diagrams
(CMDs), covering optical and near infrared (NIR) bands, and on 
updated stellar evolutionary models (\cite{gallart96a}).\par

These investigations are going to be extend out to the Local
Universe in the near future thanks to the extremely large telescopes
equipped with sophisticated Adaptive Optics (AO) systems. These new
observing facilities will allow us to explore in detail the stellar
content of elliptical galaxies, a morphological type that is not
present among Local galaxies. We need to reach a very high
spatial resolution, at the diffraction limit of 30--m class telescope,
to resolve into stars the extremely crowded regions of these galaxies
located at distances of $\sim$20 Mpc (Fornax, Virgo galaxy cluster). In the
last few years we are contributing with detailed scientific cases to
shape the technological requirements that will allow us to accomplish
the quoted scientific goals. NFIRAOS (\cite{herriot14}) and MAORY (\cite{diolaiti10,diolaiti16})
will be the first light AO modules that will assist the Thirty Meter
Telescope (TMT) and the European Extremely Large Telescope (E-ELT)
respectively. The Multi Conjugate Adaptive Optics (MCAO) modules available 
in these facilities will assure spatial resolution at the diffraction limit for a
quite large FoV (a few arcmin) when compared with Single Conjugate AO
(SCAO, tens of arcsec) modules. This will allow us to increase stellar 
statistics across the FoV, a key ingredient in resolved stellar population 
studies of nearby galaxies. Paving the way for extremely large telescopes 
means to carry out: {\it i)} systematic and detailed analyses of the performance 
expected by the MCAO modules in combination with high resolution cameras
(e.g., MAORY+MICADO\cite{davies10} system); 
{\it ii)} a comprehensive investigation of the photometric and astrometric 
performance of AO facilities currently operating at 8--10m class 
telescopes. Resolved stellar populations in Galactic stellar clusters
are the optimal benchmark to constrain these new observing
facilities. They host thousands of stars extending several arcmin on the sky, thus sampling the full
scientific FoV and they naturally provide several cluster bright stars to be used as
natural guide stars (NGS) for the tip--tilt correction.\par

A significant fraction of the 23 nights (2007--2008) that were
offered for {\it science demonstration} of the Multi--conjugate
Adaptive optics Demonstrator mounted at the VLT
(MAD@VLT\cite{marchetti08}) were mainly dedicated to stellar clusters
(\cite{bouy08,gullieuszik08,bouy09,ferraro09a,morettia09,bono10a,campbell10,crowther10,sana10,fiorentino11,ortolani11,meyer11,rochau11})
plus a few exceptions (\cite{mignani08,falomo09}). MAD (pixel
scale of 0.028 arcsec/pix) demonstrated 
the capability to correct the atmospheric turbulence effect
reaching almost the diffraction limit on a large FoV of $\sim$2 arcmin. Although, MAD was built using 
{\it ``leftovers from previous ESO (AO) projects''} (\cite{melnick12}), 
it has been a successful experiment and provided front-end science as 
clearly presented in the review paper by Melnick, Marchetti and Amico 
(\cite{melnick12}). For this reason, in 2009, the scientific community 
was quite disappointed in learning that MAD was not going to be offered 
anymore. The choice of stellar clusters as targets was mainly driven by the 
opportunity to easily identify three bright (R$\lsim$13.5mag) NGS,
within a circle of 2 arcmin of diameter, i.e., the asterism for MAD real time 
correction. It is worth mentioning that the issue of finding ``bright''
and ideal asterisms of NGS is an open problem for both
SCAO and MCAO and will only partially be solved by future all-sky surveys
(e.g. Gaia). This problem will be further enhanced when ELTs 
will become available, e.g., the limiting magnitude for NGS in MAORY is 
H$\sim$22 mag. Future ground-based optical (LSST) and space NIR 
(WFIRST, EUCLID) surveys are going to alleviate these limitations. 
Melnick, Marchetti and Amico 2012 (\cite{melnick12}) summarized the top--level
scientific requirements for a MCAO imager, they are: {\it i)} stable and
uniform PSF in a large FoV; {\it ii)} accurate photometry with a large
dynamic range; {\it iii)} high astrometric precision, all three at or
near the diffraction limit of the telescope. MAD satisfied most of
them, but there was still room for improvements in terms of PSF 
stability with time (\cite{fiorentino11}) and in terms of
astrometric precision (a factor of two larger than
expected\cite{meyer11}). The take home message from the MAD experiment 
was that classical photometric packages (like DAOPHOT\cite{stetson94}) work well with MCAO images 
as soon as the PSF is uniform. Indeed, they usally prefer uniform 
image quality to non uniform, but high--strehl ratio images.\par

The use of laser guide stars provided by the Gemini Multi-conjugate
adaptive optics System (GeMS) operating at the Gemini South telescope 
(\cite{rigaut12,rigaut14}) telescope was expected, and succeeded, to 
offer a much higher spatial uniformity and time stability for an almost
diffraction limited PSF (in K$_s$--band) 
across the $\sim$ 2 arcmin FoV. GeMS (pixel
scale of 0.02 arcsec/pix) is the only MCAO module working 
with both laser (five) and natural (three) guide stars, thus resembling 
MAORY in its MCAO mode (six laser plus three NGS). 
GeMS requires three NGS with magnitude R$\lsim$15.5 mag, 
thus increasing the MAD sky--coverage. GeMS technical performance in terms 
of Strehl ratio and Full Width High Maximum reached across the FoV, their 
dependence on the wavelength are summarized in Tables~1 and 2 of 
(\cite{neichel14a}). GeMS is routinely operating at the Gemini South 
and it already has had a strong impact on science 
(\cite{neichel14a,davidge14,saracino15,manchado15,turri15,massari16a,santos16,opitz16,bernard16}).\par

During the last few years, we have collected data of crowded Galactic
globular clusters (GCs) using the most sophisticated AO modules operating on sky, including FLAO@LBT
(\cite{esposito12}), MAD@VLT, GeMS@Gemini. We plan to
collect more data using new planned AO facilities, such as
LUCI-FLAO@LBT, Linc-NIRVANA@LBT, ERIS@VLT and HAWK-I--AOF@VLT.
Our systematic study is demonstrating that the exploitation of AO data
is far from being trivial due to the variation of the PSF across the
FoV and to its time--variability that either hamper or limit the 
co-adding of scientific frames to increase the overall Signal to 
Noise ratio. Nevertheless, these data allowed us to estimate the 
absolute ages of GCs with an unprecedented accuracy (a factor of 
two smaller than classical methods \cite{bono10a,dicecco15b,monelli15,massari16a}, see Section 2.1).
Moreover, the use of GeMS and HST images for NGC~6681 allowed us to test 
the astrometric performance that can be reached by GeMS 
highlighting the importance of accounting for geometric distortion
(\cite{massari16c}). In particular, we have shown
that it is possible to reach a precision of $\sim$0.4 mas in 
measuring the position of stars with multi epoch observations, in good
agreement with similar investigations \cite{neichel14b}. Furthermore, 
we have derived for the first time proper motions using GeMS anchored to
HST images obtaining a precision of 0.43~mas yr$^{-1}$ and an accuracy 
of 0.03~mas yr$^{-1}$  (Massari et al. 2016d, submitted). 
These errors are similar to those based only on HST images, thus
supporting current capabilities to pave the road for future developments 
in AO technology.

\section{The scientific framework: the age of the Universe and the Hubble constant H$_0$}

Cosmological results based on recent Cosmic microwave background (CMB) experiments
(Boomerang, WMAP, PLANCK), on Baryonic Acoustic oscillations
(BAO\cite{eisenstein05}), on supernovae observations (\cite{riess16,riess11a})
and on gravitational lensing (\cite{suyu13}) opened the path to the era of
precision cosmology. However, the quoted experiments are affected by an intrinsic
degeneracy in the estimate of cosmological parameters, and in particular, for the 
Hubble constant H$_0$. To overcome this problem either specific priors or the 
results of different experiments are used (\cite{bennett14}).

Recent evaluations of the H$_0$ based on CMB provide values ranging 
from 70.0$\pm$2.2 km s$^{-1}$ Mpc$^{-1}$ (WMAP9 \cite{hinshaw13}) to  
67.8$\pm$0.9 km s$^{-1}$ Mpc$^{-1}$ (\cite{planck15}). Similar values have also 
been obtained by BAO plus supernovae using the so-called inverse distance
ladder suggesting a value of 68.6$\pm$2.2 km s$^{-1}$ Mpc$^{-1}$ (\cite{cuesta16}). On the other hand, resolved objects (Cepheids plus
supernovae) provide H$_0$ values ranging from 73$\pm$2 (random) $\pm$4
(systematic)  km s$^{-1}$ Mpc$^{-1}$ (\cite{freedman10}) to 73.00$\pm$1.75 km
s$^{-1}$ Mpc$^{-1}$ (\cite{riess16}). 
Slightly larger values of the Hubble constant were obtained by using gravitational lens time 
delays (80.0$^{+4.5}_{-4.7}$  km s$^{-1}$ Mpc$^{-1}$ uniform $H_0$ in flat
$\Lambda$CDM\cite{suyu13}).\par 

The range of Hubble constant values indicates that there is some tension
between the results based on CMB and BAO and those based on primary and
secondary distance indicators. This implies an uncertainty on the age of the universe --$t_0$-- of the
order of 2 Gyr. Thus having a substantial impact not only on galaxy
formation and evolution, but also on the age of the GCs.\par  

\begin{figure*} [ht]
    \begin{center}
    \includegraphics[width=10.cm]{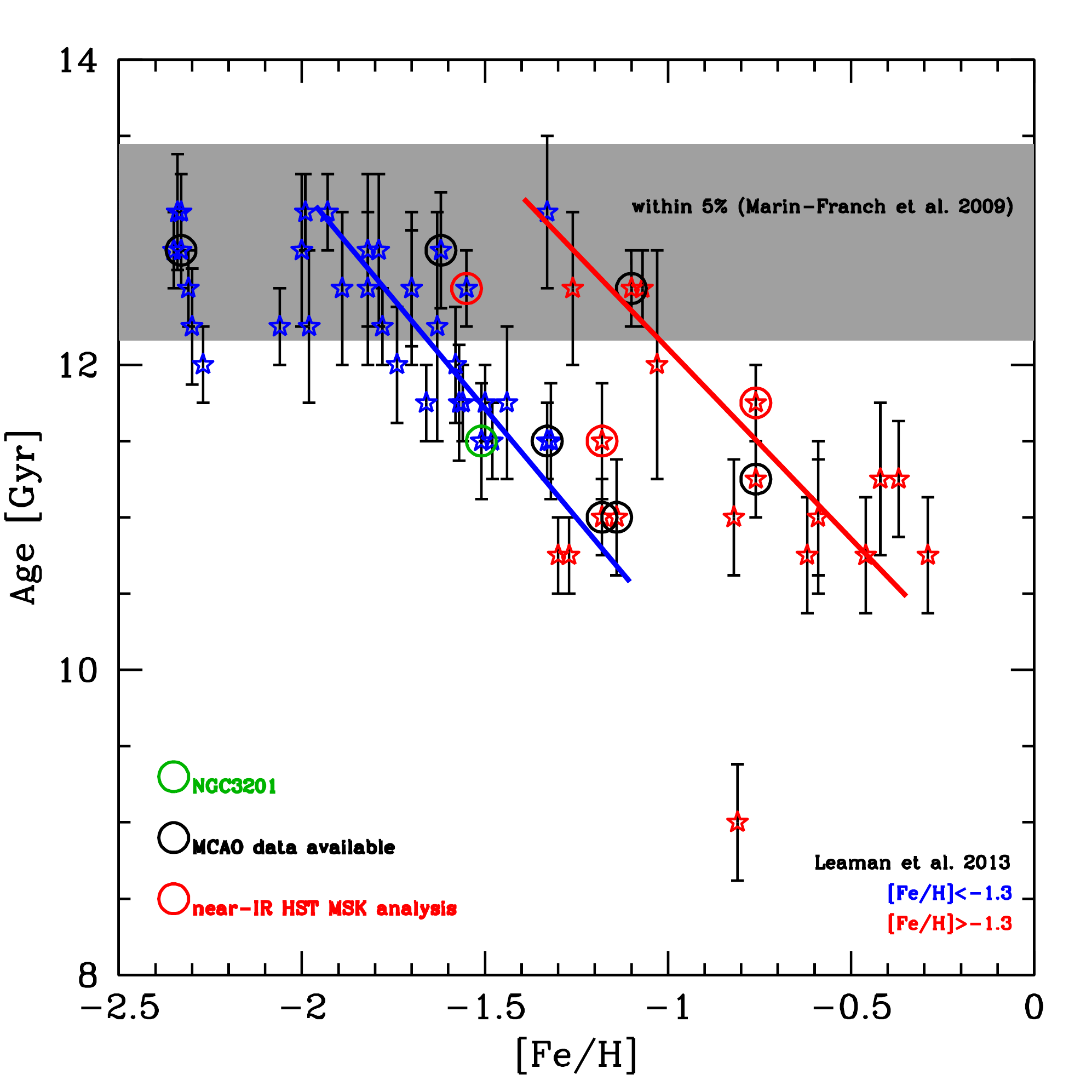}
        \caption{\small Age--metallicity diagram for GCs
          for which HST photometry is available together with the
          analysis of their relative ages. Data are taken from Leaman
          et al. 2013 \cite{leaman13}. With
          red and blue stars we have highlighted the cluster
          metallicity, i.e., metal--rich and metal--poor, respectively. Circles
          indicates the clusters for which deep MCAO (black) and HST
          (red) NIR data are available for the MSK
          characterization. Blue and red lines display the two
          age--metallicity relations \cite{leaman13}, while the grey region shows 
          the result found by Marin-Franch et al. 2009 \cite{marinfranch09}, 
          see text for more details.}
        \label{gc_age}
    \end{center}
\end{figure*}

It is worth mentioning that we are also facing a stark disagreement 
concerning the early formation of the Galactic halo. Using a sizable 
sample of GCs ($\sim$60) observed with ACS at HST, it has been suggested 
that they do obey to two different age--metallicity relations (\cite{leaman13}). 
Their working hypothesis is that GCs of the metal--rich branch were formed in 
the disk (red solid line in Fig.~\ref{gc_age}), while those belonging to 
the metal--poor one were accreted into the Halo (blue solid line in Fig.~\ref{gc_age}). 
Oddly enough, a previous study using a very similar dataset suggested 
that only a subsample of GCs do obey to an age-metallicity relation
whereas the bulk of them seems to be coeval ($\sim$ 12.8 Gyr $\pm$
5\% \cite{marinfranch09}). This suggests that the first sample was accreted from metal--poor
dwarf galaxies (blue solid line in Fig.~\ref{gc_age}) while the coeval
GCs were formed in situ (grey region in Fig.~\ref{gc_age}).
This evidence indicates that current absolute age estimates of GCs are affected 
by theoretical, empirical and intrinsic uncertainties:\\ 
{\em Theoretical}-- Stellar evolutionary models adopted to construct cluster 
isochrones are affected by uncertainties in the physical inputs. In particular, in
the adopted micro (opacity, equation of state, astrophysical screening  factors)
and in macro-physics (mixing length, mass loss, atomic diffusion, radiative
levitation, color-temperature transformations). The impact that  the quoted
ingredients have on cluster isochrones have been discussed in  detail in the
literature (\cite{vandenberg13}). The typical uncertainty in the adopted 
clock --the main sequence Turn Off (MSTO)--  is of the order of a few percent. 
Thus suggesting that theoretical uncertainties does not
appear to be the dominant source in the error budget of the absolute age of
GCs.\\
{\em Empirical}-- The main source of uncertainty in the absolute age estimate 
of GCs are the individual distances ($\Delta \mu_0 \sim$0.1 mag 
in the true distance modulus implies an uncertainty of more than 1 Gyr in 
the absolute age).\\
{\em Intrinsic}-- Dating back to more than forty years ago, spectroscopic 
investigations brought forward a significant star--to--star variation in 
C and in N among cluster stars (\cite{osborn71}). This evidence was soundly 
complemented by variation in Na, Al, and in O and by anti-correlations in CN--CH  
and in O--Na and Mg--Al (\cite{gratton12}).\\

The above evidence has further strengthened by the occurrence of multiple 
stellar populations in more massive clusters \cite{bellini13,milone14}.
However, detailed investigations  concerning the different stellar populations
indicate a difference in age that is, in canonical GCs, on average shorter than
1~Gyr \cite{ventura01}. The intrinsic uncertainty does not seem to
be the main source of the error budget of the GCs absolute age. 
To overcome the quoted uncertainties, different approaches have been suggested  
mainly based on relative age estimates, the so--called vertical and horizontal methods (\cite{marinfranch09,vandenberg13}).\par
In this context the relative age is estimated as a difference between
the clock (Main Sequence Turn Off, MSTO) and an evolved reference
point either the horizontal branch (HB) or a specific point along the
red giant branch. The key advantage of these  methods is that they are independent of uncertainties on cluster distance and 
reddening. However, they rely on the assumption that the reference points are 
independent of cluster age and introduce new theoretical uncertainties 
(conductive opacities, extra-deep mixing along the RGB). It goes without 
saying that the transformation from relative to absolute ages using a 
reference GC introduces the typical uncertainties already discussed. 
In the following we describe a new method introduced by Bono and
collaborators (\cite{bono10a}) and based mainly on deep NIR 
photometry.

\subsection{Absolute ages of Globular clusters and the Main Sequence Knee--method}

\begin{figure*} [ht]
    \begin{center}
    \includegraphics[width=9.5cm]{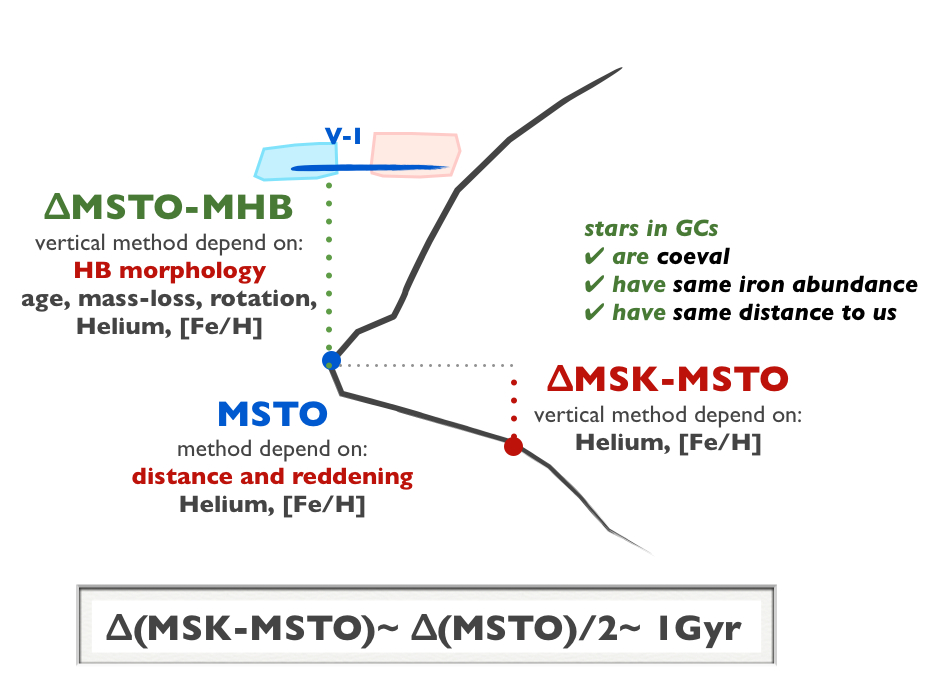}
    \vspace{+1.0cm}
        \caption{\small Cartoon representing the MSTO point in the CMD and the two vertical methods that anchor the MSTO to different 
evolved, the Horizontal Branch (darkgreen dotted vertical line), and un--evolved, the Main Sequence Knee (red dotted vertical line) 
evolutionary features. The dependencies affecting the different methods to derive absolute ages are also depicted in 
the Figure together with the improvement on the age accuracy when using the Main Sequence Knee.}
        \label{msk}
    \end{center}
\end{figure*}

The Main Sequence Knee has been observed using NIR
observations in several Galactic stellar systems, mainly globular
\cite{pulone98,pulone99,bono10a,lagioia14,milone12,milone14,turri14a,turri15,massari16a,correnti16}
and old open clusters \cite{sarajedini09a}, in the bulge
\cite{zoccali00a}. Although, this feature has been detected and
characterized in optical bands \cite{dicecco15b,monelli15}, 
it can be easily identified in the NIR regime. In cold (effective
temperatures less than $\sim$ 4000K) and low mass stars  ($\sim$ 0.5
M$_{\odot}$), the opacity in the stellar atmosphere starts to be dominated 
by the collisional induced absorption (CIA) of
transitional dipole state in molecules such as H$_2$--H$_2$,
H$_2$--He, N$_2$--N$_2$, CH$_4$--CH$_4$ \cite{saumon94,borysow97,allard95}. The CIA mechanism acts as an
absorber for wavelengths larger than 1.0 $\mu m$ and an emitter in
UV-visible light, thus causing a bluer main sequence for low mass
stars. This means that a strong change in the main sequence slope can
be observed at longer wavelengths, i.e., the Main
Sequence Knee (MSK), see Fig.~\ref{msk}.\par

Using MCAO images from MAD of the GC NGC~3201,
\cite{bono10a} suggested a new diagnostic based on the 
change in slope of the low main sequence as a precise
age--indicator\footnote{Details on the actual point used to define the
MSK is fully described in \cite{bono10a} and \cite{massari16a} and it
is the flex point before the change of curvature of the low main
sequence.}. The advantage of using this new vertical anchor when
compared to other more popular features in the CMD, such as the HB, is
its negligible age dependence. The HB instead does depend not only 
on the age but also on other unknown
parameters, i.e. mass loss, chemical composition, rotation and
everything is shaping the complex morphology of the HB
\cite{salaris16}. However, in order to reach the NIR MSK (K$_s\gsim$20 mag) in the core of
GCs from the ground we need to build deep, precise and accurate CMDs.\par 

Since the first estimate of the absolute age of NGC~3201\cite{bono10a}, several authors applied this vertical method for aging a
total of seven GCs (NGC~6838--M~71\cite{dicecco15b}; M~15--NGC~7078\cite{monelli15};
NGC~2808\cite{massari16a}; 47~Tuc, M~4--NGC~6121, NGC~2808 and
NGC~6752\cite{correnti16}). A careful analysis of the systematic and
random errors to be included in the $\Delta$MSTO--MSK method has been 
provided by (\cite{massari16a,correnti16}).

\subsection{Space motions and stellar populations with GeMS}
Our project is executed on Canadian time (PI A. McConnachie). We have
collected GSAOI data assisted with GeMS for seven Galactic GCs. They are listed in Table~\ref{tab1} and are highlighted in
Fig.~\ref{gc_age} as black circles.
The selection criteria adopted are the following: 
{\it i)} they are close enough to allow us the detection 
of the MSK (d$_{\odot} \lsim$ 12Kpc); 
{\it ii)} they suffer small amount of reddening (E(B--V)$\lsim$ 0.25); 
{\it iii)} they cover a large range in metallicity, thus providing a 
good sample to calibrate the MSK method as a function of the
metal content (-2.4$\lsim$[Fe/H]$\lsim$-0.8); 
{\it iv)} they have deep HST images that can be used as first epoch
for deriving proper motions; 
{\it v)} they span several Kpc in Galactocentric distance 
(2Kpc$\lsim$d$_G\lsim$16Kpc).

\begin{table}
\scriptsize
\caption{Properties of selected GCs for GeMS observations. NOTE: $^a$ age from Leaman et al. 2013 (\cite{leaman13}).}
\label{tab1}
\begin{center}
\begin{tabular}{lcccr}
\hline
\hline
GC's name    & $<$[Fe/H]$>$     & age$^a \pm \sigma_{age}$  & d$_G$  & M$_V$\\
             &                  &   Gyr                      & Kpc    &  mag \\
\hline
NGC1851	        &  -1.18&	11.00 $\pm$ 0.25 &		16.6&	-8.33 \\   	    
NGC2808	        &  -1.14&	11.00 $\pm$ 0.38 &   	11.1&	-9.39 \\      
NGC5904(M5)	&  -1.33&	11.50 $\pm$ 0.25 &   	6.2 &   -8.81 \\  
NGC6681(M70)    &  -1.62&	12.75 $\pm$ 0.38 &   	2.2 &   -7.12 \\  
NGC7078(M15)    &  -2.33&	12.75 $\pm$ 0.25 &   	10.4&	-9.19 \\  
NGC6723	        &  -1.10&	12.50 $\pm$ 0.25 &   	2.6 &   -7.83 \\      
NGC6652	        &  -0.76&	11.25 $\pm$ 0.25 &   	2.7 &   -6.66 \\	    
\hline
\end{tabular}
\end{center}
\end{table}

Our sample (and in particular the age of NGC~6652), when fully analysed,
will provide solid constraints on the two proposed Galactic halo
formation scenarios (Fig.~\ref{gc_age}, \cite{leaman13,marinfranch09}). So far we have carefully
analysed data for NGC~1851 (\cite{turri14a,turri15}) and NGC2808
(\cite{massari16a}) and we have obtained accurate and precise CMDs
reaching K$_s\sim$ 21.5 mag, thus well below the MSK point. However,
firm conclusions require more data of metal--rich GCs belonging to the
{\it young} branch (age smaller than 11.5~Gyr, see Fig.~\ref{gc_age}): i.e.,
NGC~5927, NGC~6304, NGC~6352, NGC~6366, NGC~6496, NGC~6624, NGC~6637
(M~69) and Pal~12. Among the more metal--rich GCs, the MSK method has been applied to 47Tuc
\cite{correnti16} and M~71 \cite{dicecco15b} and it returns age values
compatible within one $\sigma$ to that obtained by Vandenberg and collaborators
(\cite{vandenberg13,leaman13}), i.e., 11.75 vs 11.6 Gyr and 11.00
vs 12.00 Gyr, respectively. For NGC~6652, the data are already in hand but we need
Flamingos--2 calibration data (accepted with priority B) in order to
complete our analysis.

\section{The impact of precise and accurate photometry on resolved stellar populations}

The difference between photometric precision and accuracy can be
synthesized in our ability to repeat photometric
measurements around a specific mean value (with a small dispersion) and
in our ability to measure the {\it true} magnitude of a star. We discuss 
more in detail the impact that these two concepts have on the construction 
of CMDs.
In a precise and accurate CMD we require evolutionary sequences to
appear well defined within a few hundredths of a magnitude. Once properly
modeled these CMDs allow us a complete characterization of the global
properties of the stellar system, and in the understanding of 
their formation and evolution (\cite{gallart96a}). Photometric precision and accuracy are 
tightly coupled issues in dealing with ground--based and space observations. 
The former issue is mainly related to how well we model the PSF across 
the scientific FoV; special attention is payed to crowded stellar 
environments (such as cores of GCs). The latter issue is linked to our 
knowledge of the observing facility (optics plus camera) and on the
availability of a sizable sample of local stellar standards within the
scientific FoV. The absolute photometric calibration might include 
several ingredients: magnitude and color terms, atmospheric-extinction 
and spatial dependence.\par 

When dealing with AO observations (and in particular with SCAO), the complexity in 
modeling the spatial variation of the PSF and the limited FoV (small sample of 
calibrating stars) makes more difficult the effort to obtain accurate and 
precise CMDs. The former problem can be mitigated whenever a large number of 
bright and isolated sources are present in the field and a detailed approach in 
modeling the PSF is followed (\cite{schreiber11,fiorentino11,schreiber13}). 
Using the NIR images collected for the GC M15 with PISCES--FLAO@LBT, we found 
that a careful data reduction has to be performed in dealing with images 
from SCAO modules. This limitation becomes even more severe when exploring
the core of GCs (\cite{fiorentino14b}). We have performed photometry
of these images using an updated version of ROMAFOT that offers the
opportunity to model the PSF variation across the FoV with asymmetric 
analytical functions. The impact of 
asymmetric PSF on the CMD of M15 is a well defined MSTO and narrower MS, RGB 
and HB sequences (see Fig.~6 in \cite{fiorentino14b}). The width in colour of the quoted evolutionary features is a 
solid science--based approach to quantify photometric errors. In principle
photometric errors are the main cause of the broadening of the CMD features, 
since the underlying stellar population is almost coeval and chemically homogeneous. 
Furthermore, a large number of calibrating standard stars in the field is 
mandatory to obtain an accurate photometric calibration, and to check if residual positional effects
are present, due to the variation of the PSF.\par

Although, in the case of MCAO the above problems are naturally
alleviated, they are still worrisome. We have presented the case of
NGC~1851 observed with GeMS (\cite{turri15}) where a
fine--tuning in the PSF modeling and in the (absolute and inter-chips)
calibration enabled us to disentangle the two sub--giant branches
previously observed with HST in this cluster. The offset between the two sequences
is $\delta_{OFFSET}\sim$ 0.05 mag at V$=$19mag. These double sub--giants
may represent an age separation of $\sim$1~Gyr or
a chemical difference between two coexisting populations in
NGC~1851 \cite{milone08}. Evolutionary models, supported 
by detailed spectroscopic observations of Red
Giant Branch stars, seem to prefer the latter hypothesis. They suggest that
the bright sequence is associated to a stellar population with a normal 
$\alpha$ enhancement while the faint one to another population with an
overabundance of C+N+O (\cite{cassisi08}). A more detailed
discussion of the full data reduction using  
DAOPHOT suite of programs and
several technical issues can be found in Turri et al. 2016 (this
conference \cite{turri16}). Here, we remember that DAOPHOT describes the PSF model
by the sum of a symmetric analytic bivariate function (typically a
Lorentzian or Moffat) and an empirical look--up table representing corrections
to this analytic function from the observed brightness values
within the average profile of several bright stars in the image.
The empirical look--up table makes it possible to account for the
PSF variations (from linear to cubic) across the FoV.
In Turri et al. 2016\cite{turri16}, we show how an increase in the
degree of the PSF variation across the FoV is needed in order to
minimize the residuals left over by the subtraction of all the stars
from GeMS images using the assumed PSF model. This is also visible by
an inspection of the CMDs obtained using different
assumptions on the PSF variation. Furthermore, we have
also shown that the choice of the PSF radius has a strong impact on
the width of the main evolutionary sequences.

\begin{figure*} [ht]
    \begin{center}
    \includegraphics[width=8.cm]{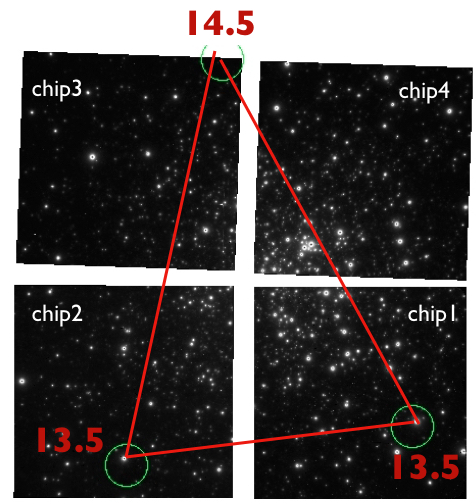}
    \includegraphics[width=8.cm]{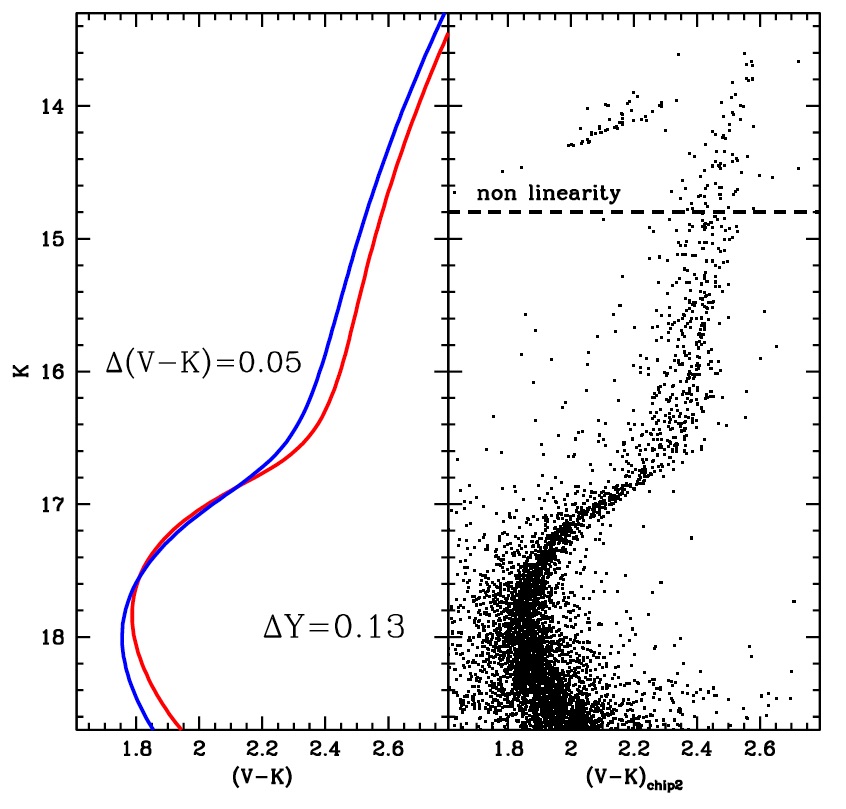}
        \caption{\small {\it Left panel:} Image for NGC2808 taken
          with GSAOI@GeMS. The asterism of the NGS used is highlighted
          together with their R-band magnitudes. {\it Right panels:} {\bf sx--} Theoretical isochrones for
          the metallicity of NGC~2808 and for two helium abundances,
          primordial He (Y$=$0.245, red line) and enriched of $\Delta
          Y=$0.13 (blue line); {\bf dx--} Optical vs NIR CMD for 
          NGC~2808 obtained using only stars located in Chip2 of GeMS images. 
          The non linear regime has been highlighted with a dashed line.}
        \label{NGC2808_He}
    \end{center}
\end{figure*}

We have performed a similar photometric analysis using GeMS data for
NGC~2808 (fully described in \cite{massari16a}). An example of our 
images is shown in Fig.~\ref{NGC2808_He} (left panel) together with 
the asterism used for the NGS. This GC has the same metallicity of 
NGC~1851, but different HB morphology.
NGC~2808 hosts several (up to five) 
stellar sub--populations that could be modeled with different helium 
abundances ($\Delta Y=$0.13 \cite{milone15}). The complexity of NGC~2808 is
even enhanced by the discovery of five distinct groups of stars along
the Na-O anti--correlation \cite{carretta15} not exactly corresponding 
to those identified using different helium abundances (\cite{milone15}). 
Fig.~\ref{NGC2808_He} (right--sx panel) shows the expected maximum 
split in colour of the RGB due to $\Delta Y=$0.13 is V$-$K$_s\sim$ 0.05 mag at K$_s\sim$16
mag. The optical--NIR CMD derived combining HST and GeMS data show a
significant spread in the RGB not present in the SGB (right--dx panel). 
Although, we are not able to identify sub--structures in the RGB, its 
width in colour seems to be real. Using the images collected in eight 
different epochs we estimated the precision (repeatability) of our measurements that is
$\sigma$(K$_s$)$\sim$ 0.03 mag. Thus supporting the lack of a clear color separation 
among the different sub--populations. In passing we note that the combination 
of optical (V) and NIR (K$_s$) bands it is not optimal for the study 
of multiple helium sequences. A complex helium abundance distribution,
as that suggested for NGC~2808, would require a precision
smaller than 0.01 mag to be fully detected.\par

\begin{figure*} [ht]
    \begin{center}
     \vspace{-2cm}
    \includegraphics[width=8.cm]{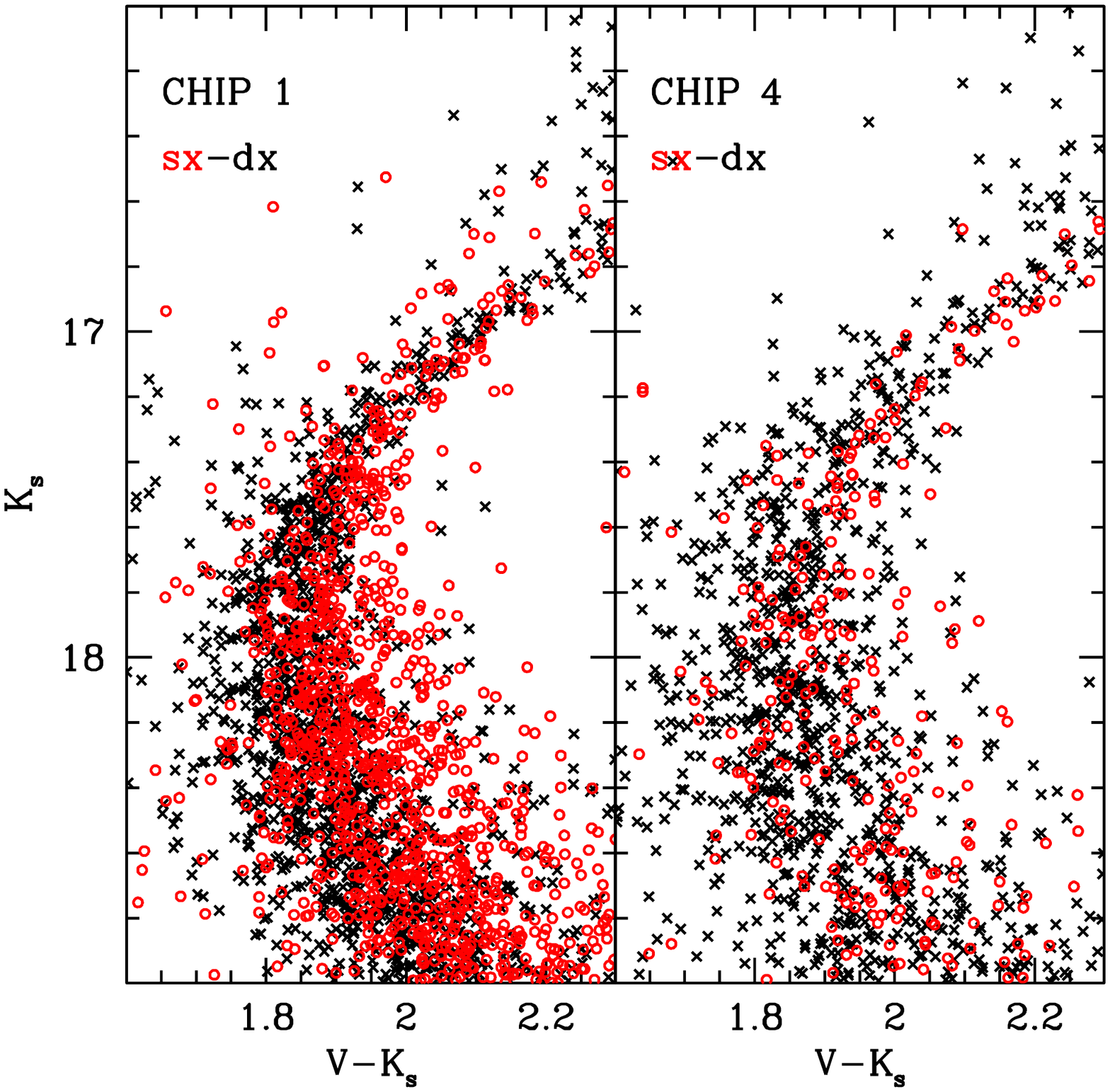}
    \includegraphics[width=8.cm]{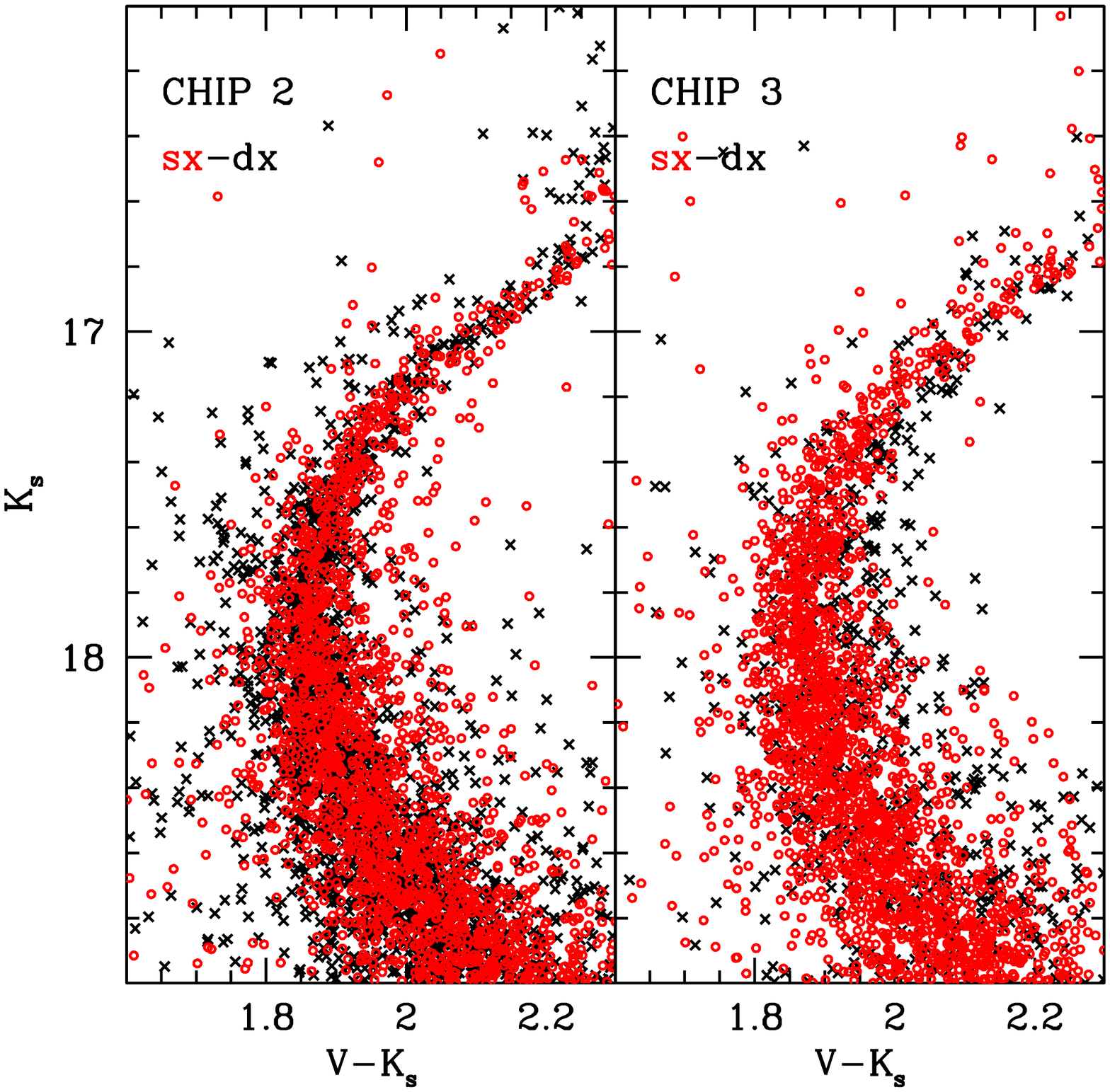}
    \vspace{-2.cm}
        \caption{\small From left to right panels, CMDs obtained for the four chips of GeMS, the number is labeled in each panel. We have used different colour--code in order to highlight photometry coming from the left (sx, red) and right (dx, black) regions on individual chips.}
        \label{NGC2808_cmd}
    \end{center}
\end{figure*}

\begin{figure*} [ht]
    \begin{center}
     \vspace{-2cm}
    \includegraphics[width=8.cm]{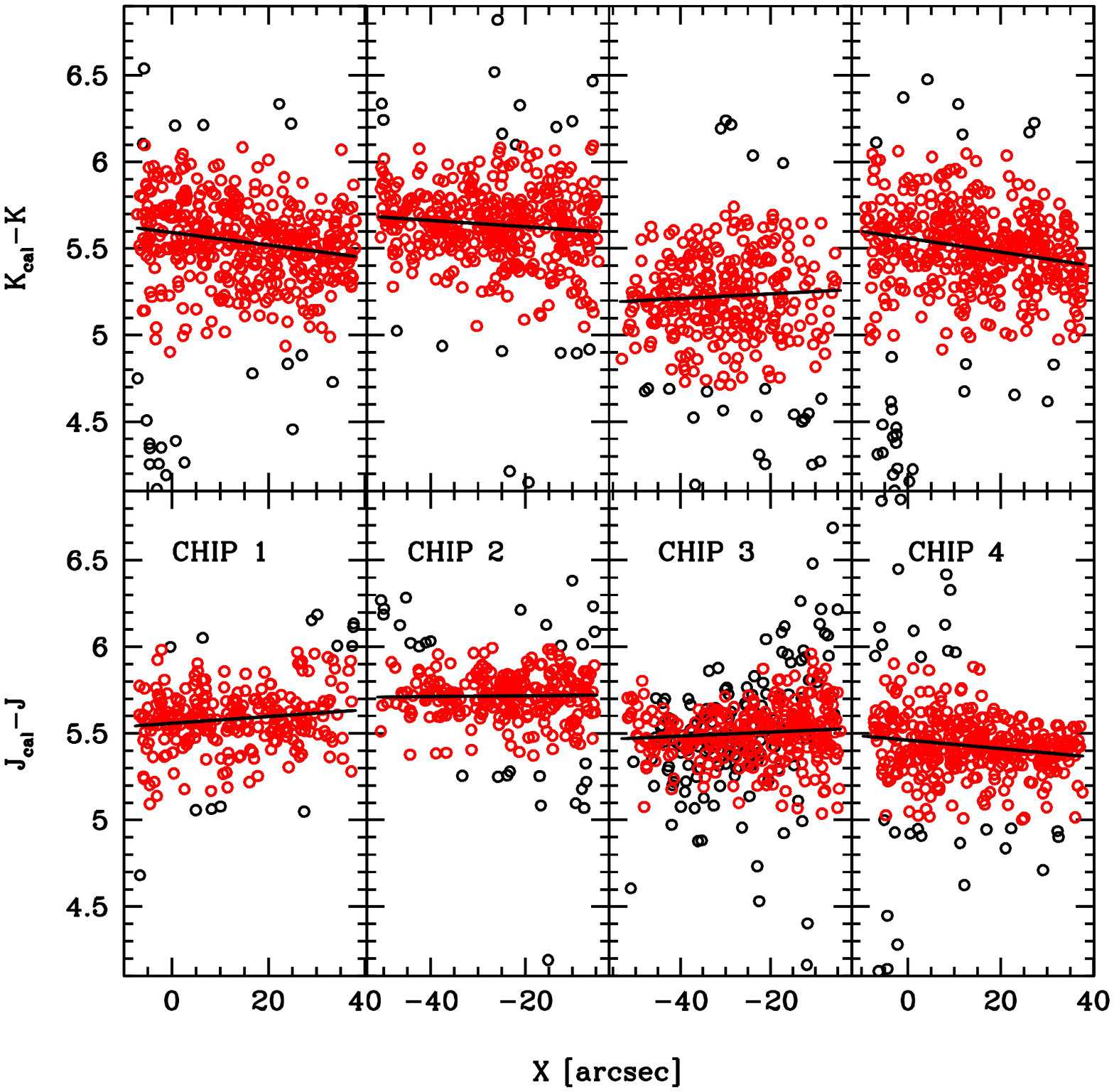}
    \includegraphics[width=8.cm]{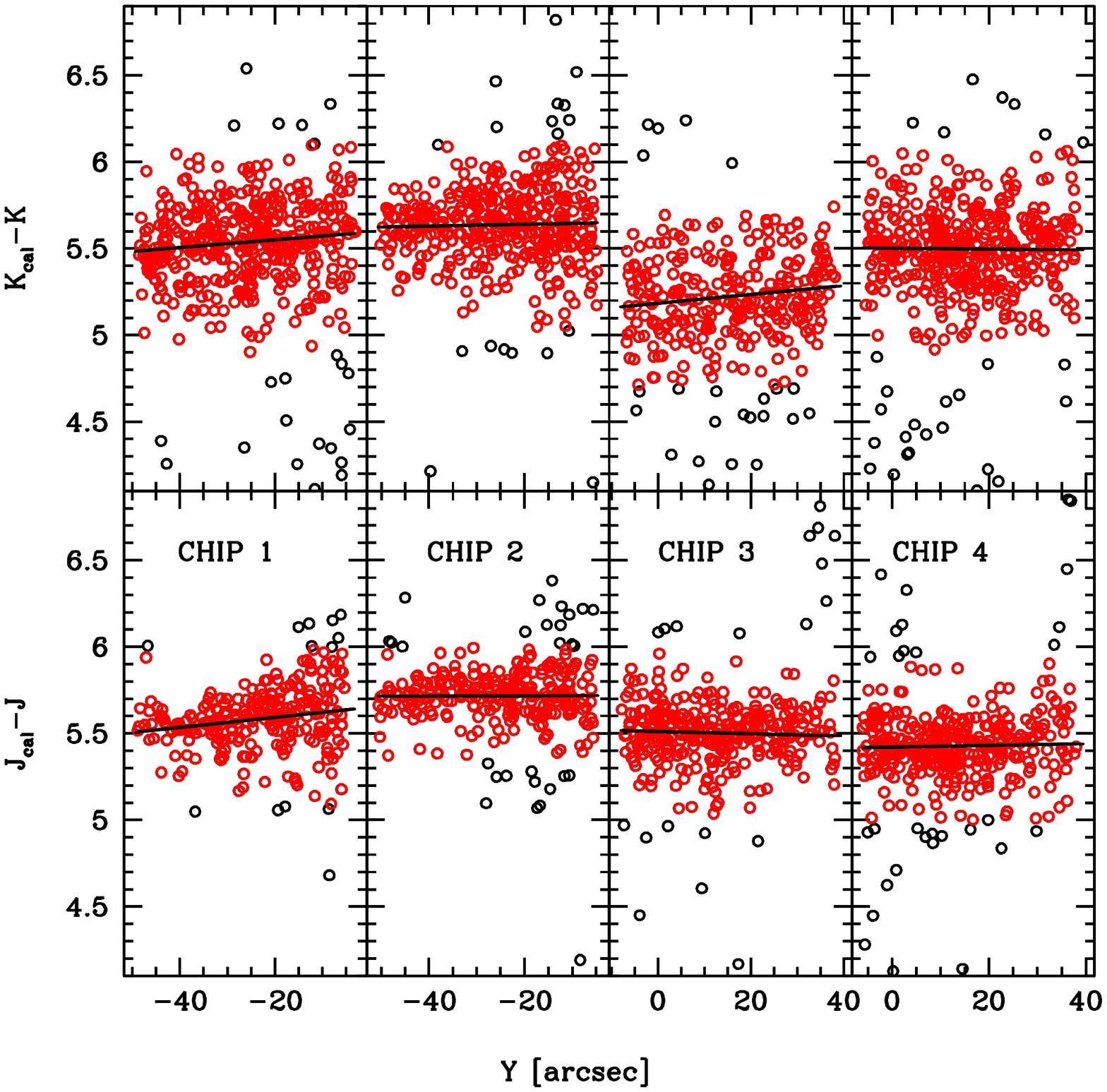}
    \vspace{-2.cm}
        \caption{\small {\it Left panel:} From the left to the right Chip 1 to Chip 4. X-axis spatial distribution of the difference in magnitude between GeMS and HAWK-I calibrating stars (red open circles) in both K and J filters. Black solid lines show the linear fit. {\it Right panel:} Same the left panel for the Y-axis.}
        \label{NGC2808_cal}
    \end{center}
\end{figure*}

We have performed the calibration onto the 2MASS photometric system by
means of an intermediate catalogue derived using archival HAWK-I data
for NGC~2808. Mounted at UT--4, HAWK-I camera has the advantage to
have a large FoV ($\sim$ 7.5$\times$7.5 arcmin) and a small pixel
scale (0.106 arcsec/pixel). Thus it offers the maximum overlap with
both the deep diffraction limited GeMS data (14 mag$\lsim$ K $\lsim$21.5 mag) and the shallow 2MASS seeing--limited data (K$\lsim$14 mag), thus improving the final photometric accuracy.
We have selected a range of about 2 magnitudes (15 mag$\lsim$ K
$\lsim$17 mag) that do not suffer non--linearity in order to make a
reliable calibration of each GeMS chip to HAWK-I data. Although, we
have applied only zero points (no trend with magnitudes and colours
was found) and we have used the same approach in calibrating the four
chips, the final result does not seem satisfactory neither for the
inter--chips calibration\cite{massari16a} nor for the existence of
spatial trend within each individual chip, as shown in
Fig.~\ref{NGC2808_cmd}. We have divided the single--chip photometry in
two catalogues along the x--axis. Stars on the left side of the chip
(sx--stars) are indicated with red open circles and those on the right
side (dx) with black crosses. By an inspection of this figure, it is
clear that the mean position of the MSTO show a clear x--axis trend in
three chips out of four. The only exception is Chip 2. This may depend
on the major uniformity of the MCAO correction, i.e., the
FWHM is $\sim$0.08 across the full Chip 2 as shown in Fig.~1 (left
panel) of Massari et al. 2016 (\cite{massari16a}). This allows us a
better modeling of the PSF across the Chip 2. Only the photometry of
Chip 2 was used for the estimation of the absolute cluster age with 
the MSK method. \par 

Although, seeing--limited data strongly suffer of blending effect
(\cite{turri16}), we can use them to test the spatial dependency of
our photometry across the FoV, see Fig.~\ref{NGC2808_cal}. In this
figure differences between GeMS and HAWK-I magnitudes for calibrating
stars are shown as a function of X and Y axes for both J and K
bands. The linear fit for the stars are
also shown (black solid lines), their slopes range from 0.0002--0.002
for the J--band and from -0.001--0.004 for the K--band in the X-direction 
and from -0.0006--0.003 for the J--band and from -0.0002--0.002 for 
the K--band in the Y-direction. Chip 2 again shows the smallest spatial dependence when
compared with the others, even thought the slope of the K--band in 
the X--direction is not zero (-0.0018) we neglected the correction because the 
magnitude difference distribution shows a large scatter ($\sigma/$N$_{stars} \sim$0.01 mag).
In a forthcoming investigation we plan to better model this spatial variation in
order to use the photometry of the four chips to build a precise and accurate
CMD of NGC~2808.

\subsection{Numerical experiments on crowding and photometric errors}

\begin{figure*} [ht]
    \begin{center}
    \includegraphics[width=8.cm]{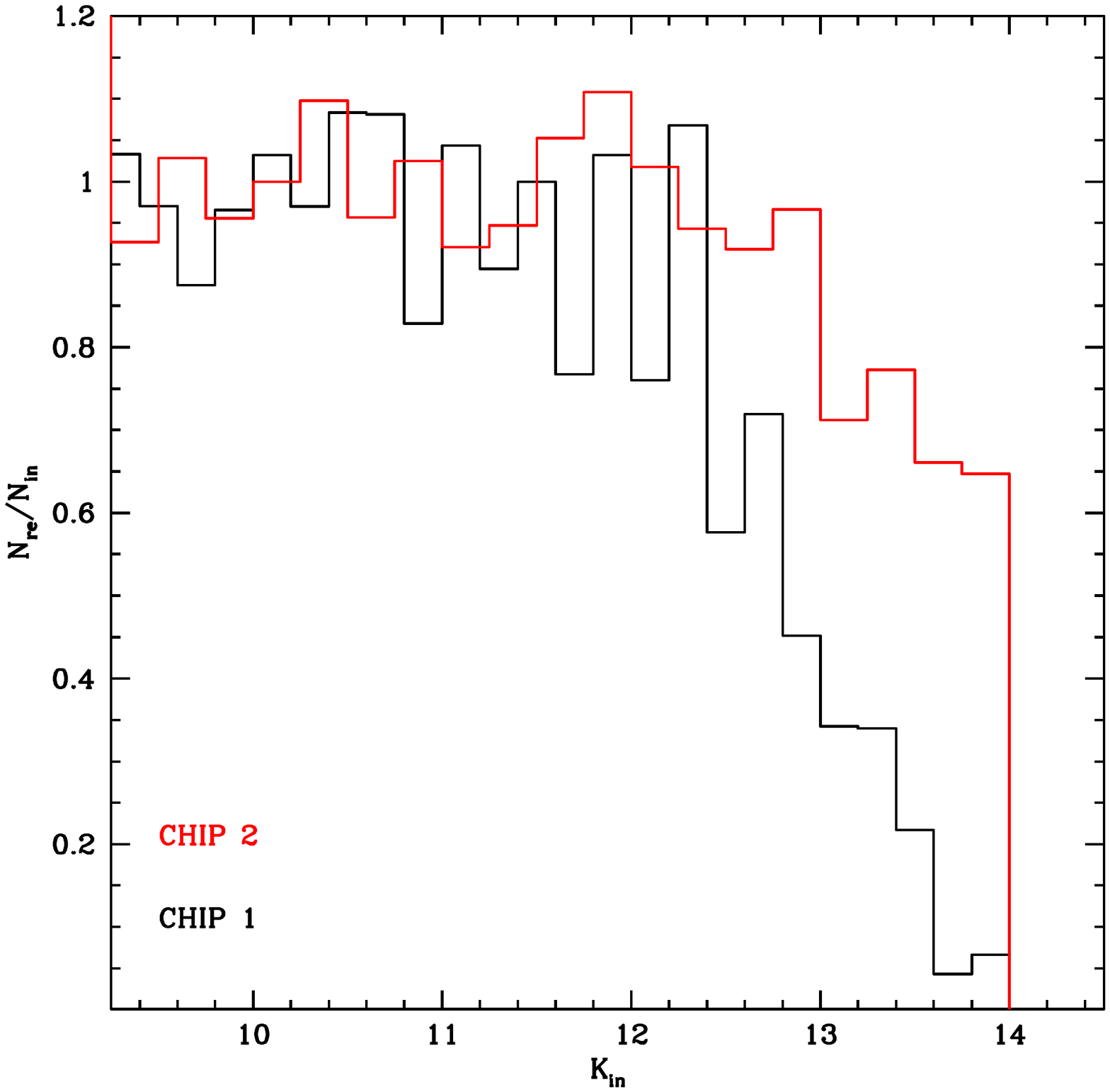}
    \includegraphics[width=8.cm]{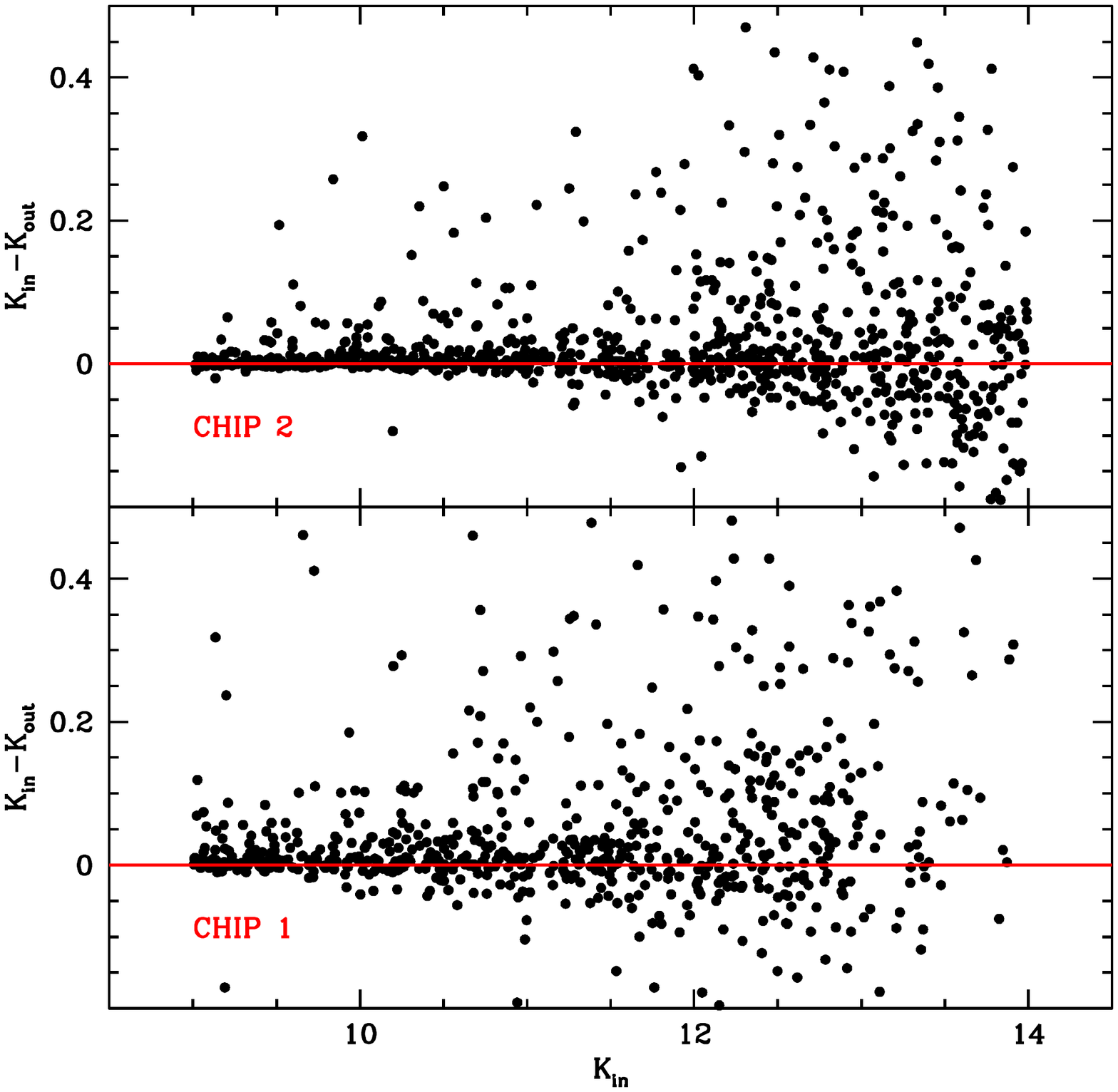}
    \vspace{-2.cm}
        \caption{\small {\it Left panel: The ratio between the number of retrieved
and injected stars when performing an artificial star test. The same ratio is also called {\it
completeness}. 
        This test was applied to one K$_s$-band image of NGC~2808: chips 1 and 2.} . 
        {\it Right panel:} Difference in magnitude between injected and retrieved
stars adopted 
        for the artificial star test.}
        \label{NGC2808_art1}
    \end{center}
\end{figure*}

In order to quantify photometric errors in the constructed CMDs, one
of the most popular approach is to perform an artificial star test. This
method consists in adding a certain number of artificial stars of known
magnitude (injected synthetic stars) to real images and then
in extracting their magnitudes by performing once again the photometry 
(see for more details \cite{gallart96a,gallart96b}). This step is 
fundamental in the comparison between observations and theory, and in turn 
in the correct interpretation of the CMD. DAOPHOT suite of programs has a sub-routine
(ADDSTAR algorithm \cite{stetson94}) to inject artificial stars given
a grid of magnitudes and positions. The comparison between injected and retrieved 
stellar magnitudes allow us to quantify the completeness and the photometric errors 
for each bin of magnitude. The full procedure
is based on the assumption that we have a good model of the PSF that 
can allow us to add stars in real 
images. This is a solid assumption for space--based observations for which we have
an accurate knowledge of the final PSF, since it is stable in space and in time. 
However, this is not the case for ground--based images, in particular, for those 
taken with AO--modules. As a matter of fact, the results of this test are expected 
to be strongly related to our ability in modeling the PSF across the
FoV. This is the reason why artificial star tests have rarely been applied 
to AO--images. The most popular approach to quantify the photometric completeness
of images collected with AO--modules has been the use of deep HST photometry 
(\cite{fiorentino11,massari16a}).\par

Here we present a simplified test for one image in K$_s$--band of the
GC NGC~2808. K$_s$ band is expected to have the largest spatial
dependence as shown in Fig.~\ref{NGC2808_cal}. In particular, we focus 
our attention on the two chips showing the smallest (Chip 2) and the 
largest (Chip 1) spatial colour variation along the X-direction (see 
Fig.~\ref{NGC2808_cmd}). The PSF was modeled using DAOPHOT: 
{\it i)} we selected $\sim$100--200 bright and isolated stars uniformly 
distributed across the FoV; 
{\it ii)} we fit their brightness profile with an analytic function plus 
a look--up table of the residuals that vary 
across the FoV with a cubic dependence on the position; 
{\it iii)} we injected stars with instrumental magnitudes ranging from K$_s
\sim$ 9 to 14 mag accordingly to this PSF model. This interval
corresponds to calibrated K$_s$ magnitudes from $\sim$14.5 to 19.5
mag. 

Fig.~\ref{NGC2808_art1} (left panel) shows that the completeness 
is dropping down very rapidly in Chip 1 when compared
with Chip2, similarly to what found using HST photometry (see Fig.~5
of \cite{massari16a}). Moreover, the photometric errors
(Fig.~\ref{NGC2808_art1}, right panel) also show a relevant decrease
in the MCAO performance between the two chips. The difference 
K$_{in}$-K$_{out}$ is less than $\sim$0.01 mag down to $\sim$12 mag 
and to $\sim$14 mag in Chips 1 and 2, respectively. The dispersion 
around the mean value of the magnitude difference is also a good
approximation of the photometric error (\cite{deep11}). It typically 
ranges---when moving from bright (K$_s\sim$ 9 mag) to faint 
(K$_s\sim$ 14 mag)---from 0.03 to $\sim$0.15 mag for Chip 1 and 
from 0.005 to $\sim$0.1 mag for Chip 2.\par

\begin{figure*} [ht]
    \begin{center}
    \includegraphics[width=8.cm]{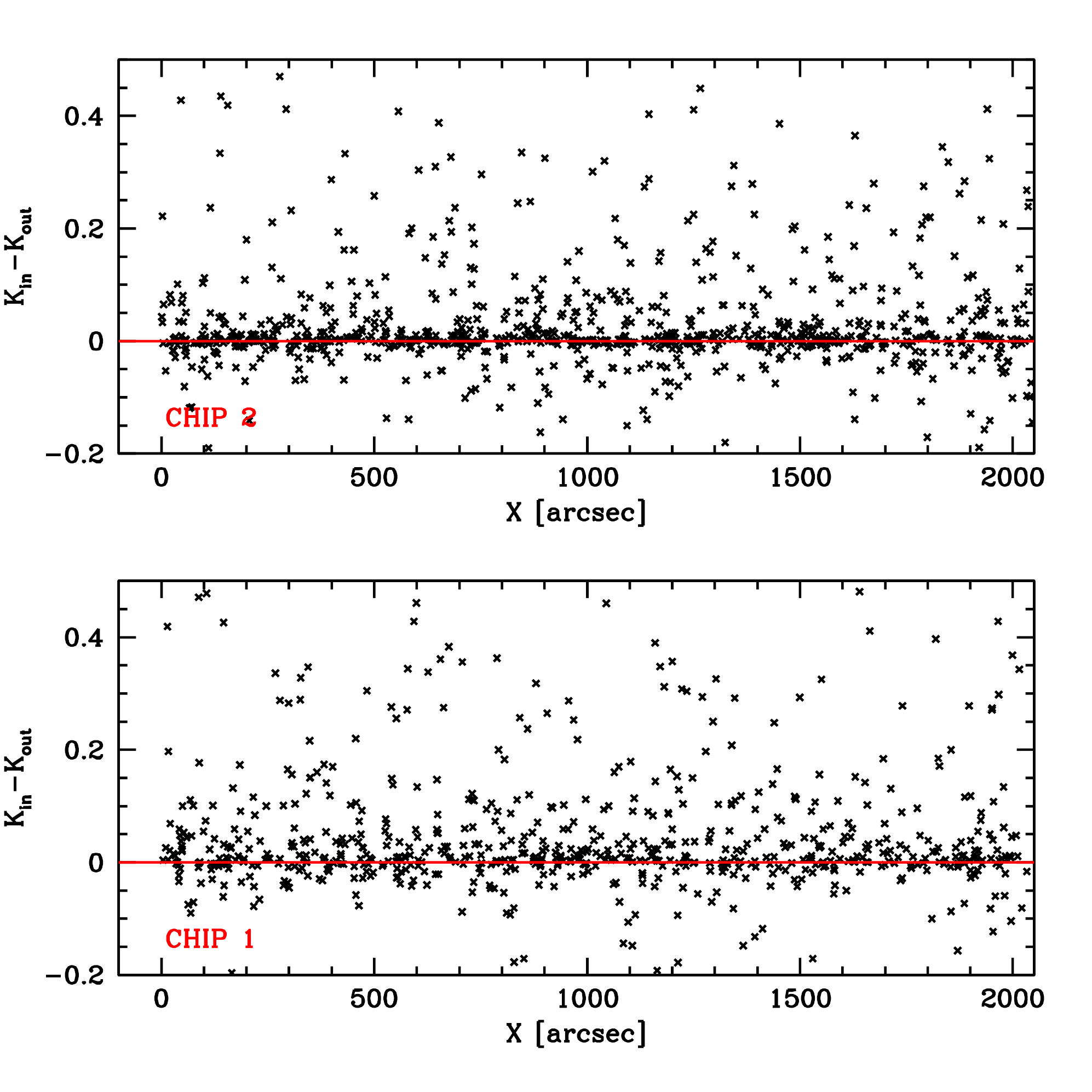}
    \includegraphics[width=8.cm]{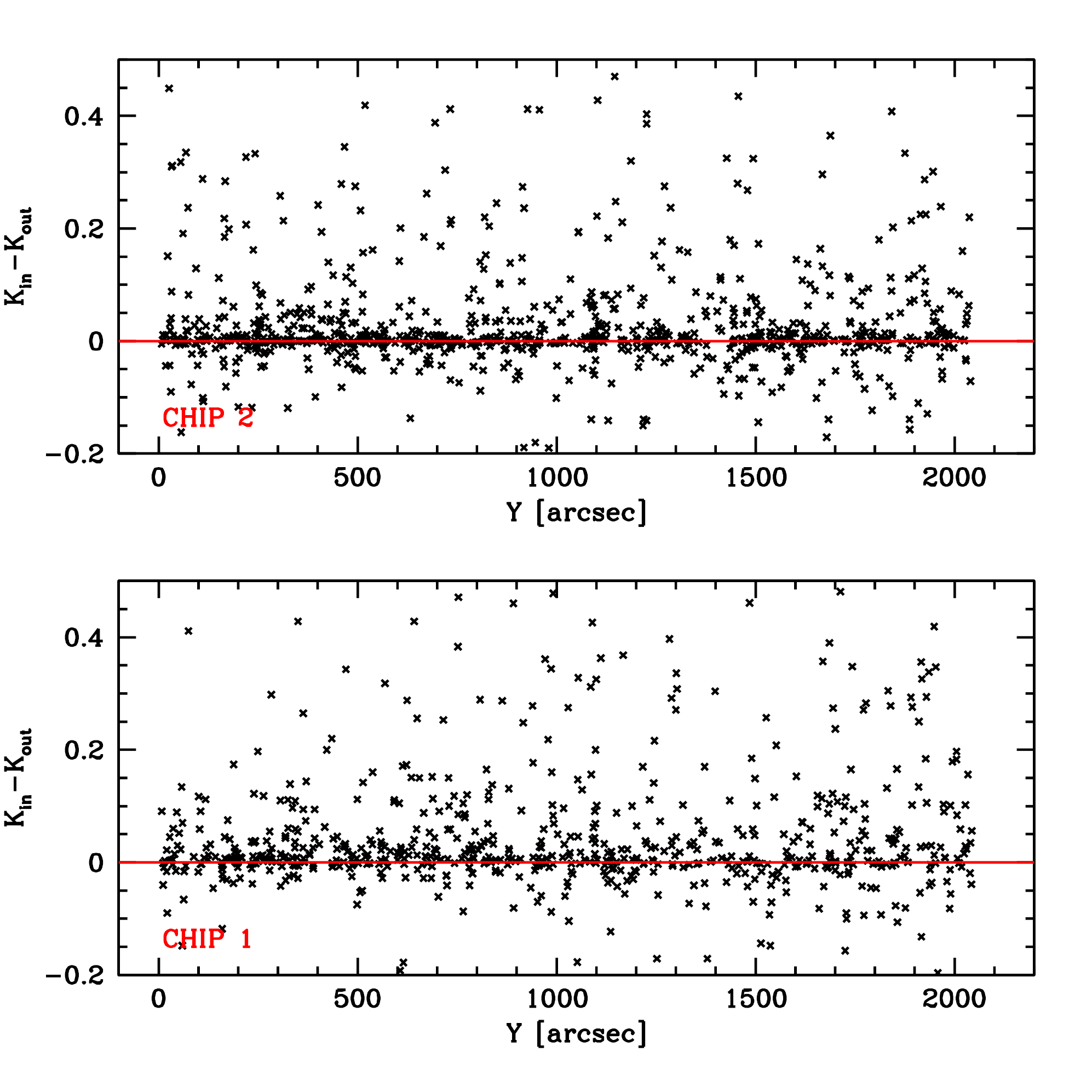}
        \caption{\small Spatial variation along the X (left) and the Y (right) 
         direction of the injected and retrieved magnitudes for the 
         artificial star test of Chip 1 (bottom panels) and Chip 2 (top panels).}
        \label{NGC2808_art2}
    \end{center}
\end{figure*}

Finally, we investigated the behaviour of the magnitude difference between
injected and retrieved stars as a function of the X and the Y--direction,
(see Fig~\ref{NGC2808_art2}). A glance at the content of this figure reveals 
that there is no trend with the position of the stars, this applies to both 
Chip 1 and Chip 2. However, we have already discussed in the previous section 
that real  seeing limited stars--observed with HAWK-I at VLT--- display a 
well defined radial trend in the X--direction of Chip 1. This evidence seems 
to suggest that a classical artificial star test applied to AO-images may 
underestimate photometric errors when a variation of the PSF
across the FoV can not be neglected. This is the consequence of the poor a--priori knowledge 
of {\it true PSF} model, since we are forced to use the {\it extracted
  PSF} to inject stars into the real image. This notwithstanding the artificial 
star test seems to be appropriate to quantify the degree of completeness of the 
stellar populations and to estimate how the crowding affects the photometry.

\section{Summary and conclusions}

During the last few years we are experimenting a new class of AO
facilities at the 8--10 m class telescopes, e.g., MAD@VLT, MAD@VLT, 
GeMS@Gemini, PISCES and LUCI-FLAO@LBT, Linc-NIRVANA@LBT, ERIS and 
HAWK-I--AOF@VLT. In this investigation we have focussed our attention to 
MCAO systems that provide a quite uniform correction on a moderately 
large FoV (a few arcmin). These features are ideal to study in detail 
resolved stellar populations in the Milky Way (GCs, Galactic bulge, Nuclear Bulge) 
and in nearby galaxies. Detailed resolved stellar population studies allow us also
to characterize the astrometric and photometric performance of these
sophisticated AO systems. This is a nowadays stepping stone to fully exploit 
the unprecedented observing facilities of near future ELTs.\par
We have focussed our attention on data taken with 
GeMS@Gemini--South telescope for some Galactic GCs. 
They have been already presented in a series of papers
(\cite{turri14a,turri15,turri16,massari16a} Massari et al., 2016,
A\&A, submitted). We have carefully analysed these images reaching a
limiting magnitude in K$_s\sim$ 21.5 mag and the main scientific
outcomes are the following: {\it i)} disentangle the two populations
showing up along the SGB of NGC~1851 and detect the MSK; {\it ii)} accurate and precise estimate of the absolute age 
of NGC~2808, t$=$10.9$\pm$0.7 (intrinsic)$\pm$0.45
(metallicity term) Gyr, in good agreement with previous
studies; {\it iii)} to derive precise (0.43~mas yr$^{-1}$) and accurate (0.03~mas yr$^{-1}$) 
proper motions for NGC~6681 using HST data as first epoch and GeMS as
second epoch.\par
The overall scientific analyses show that the quality of GeMS
data is enough to reach astrometric and photometric precisions similar
to that of HST in optical bands. In this investigation we also outline 
the main issues when dealing with AO images and the need for a new genuine
approach. The leading steps in this novel approach are the following:\par  
{\bf PSF modeling:} Although, MCAO corrects a quite large FoV
  and provide an almost uniform diffracted limited PSF across the images, the residual 
  variation of the PSf can not be neglected and has to be carefully 
  modeled. We have shown that standard photometric data reduction on 
  MCAO images is feasible. However, there are still some residual 
  effects that need to be undertaken to optimize the scientific return 
  of these new challenging instruments. This requirement becomes even 
  more relevant in dealing with SCAO images. In the last years, the 
  community devoted a paramount effort in updating existing and in 
  developing new software to accomplish data reduction of AO images: 
  {\it STARFINDER}, with an hybrid modeling that allows the adaptation of the
  parameters of the analytical part over the FoV and also takes into
  account for the contribution of the numerical residuals
(\cite{diolaiti00,schreiber11}); {\it PATCH}, a software developed for image restoration with spatially variable 
  PSF (\cite{lacamera15}); {\it ROMAFOT} with an analytical PSF model
  that varies across the FoV and account also for asymmetric
  components (\cite{buonanno89,fiorentino14b}). It is worth to mention that
  information coming from PSF--reconstruction techniques
  \cite{jolissant14} can provide a robust ingredient for these update
  versions of photometric data reduction codes.\par 

{\bf Complex calibration:}  To obtain accurate and
  precise CMDs we need a careful absolute calibration. All sky NIR survey performed by 2MASS was a quantum 
jump in photometric accuracy, since it opened a new and solid approach to calibrate 
NIR photometry. However, it is too shallow (K$\lsim$14,H$\lsim$15,J$\lsim$15.8 mag) 
and the spatial resolution ($\sim$ 0.5 arcsec) inadequate for crowded
stellar fields. Indeed the large pixel scale makes difficult to select not blended
  stars in 2MASS catalogues and faint 2MASS stars are typically
  saturated in images collected with 8--10m class telescopes assisted with AO. An intermediate 
step is necessary and a possible strategy might be the pre--AO--imaging
with similar 8--10m telescopes, but either in seeing limited conditions
(e.g., Flamingos--2 at Gemini South) or enhanced by GLAO systems
(e.g., HAWK--I at VLT). Once again the smaller FoV offered by SCAO systems makes these issues even more complex when compared with
  MCAO images. As extensively discussed here, the opportunity to have at disposal
  such intermediate catalogues allow us to investigate in detail the
  technical performance of AO--systems. These NIR calibrating stars
  will also be crucial in the future to select the NGS and in turn the optimal
asterism for ELTs.

\acknowledgments 
Based on published results obtained using MAD at the VLT and on proprietary observations obtained at the Gemini Observatory and acquired through the Gemini Science Archive.
GF and DM has been supported by the FIRB 2013 (MIUR grant RBFR13J716). GF thanks the SPIE organizers for this invited paper and B. Neichel and G. Sivo for the useful discussion during this 2016 SPIE conference in Edinburgh.

\bibliography{ms} 

\begin{thebibliography}{10}

\bibitem{gallart96a}
{Gallart}, C., {Aparicio}, A., and {Vilchez}, J.~M., ``{The Local Group Dwarf
  Irregular Galaxy NGC 6822.I.The Stellar Content},'' {\em \aj}~{\bf 112},
  1928--+ (Nov. 1996).

\bibitem{herriot14}
{Herriot}, G., {Andersen}, D., {Atwood}, J., {Boyer}, C., {Byrnes}, P.,
  {Caputa}, K., {Ellerbroek}, B., {Gilles}, L., {Hill}, A., {Ljusic}, Z.,
  {Pazder}, J., {Rosensteiner}, M., {Smith}, M., {Spano}, P., {Szeto}, K.,
  {V{\'e}ran}, J.-P., {Wevers}, I., {Wang}, L., and {Wooff}, R., ``{NFIRAOS:
  first facility AO system for the Thirty Meter Telescope},'' in [{\em Adaptive
  Optics Systems IV}{\nolinebreak\hspace{0.1em}]},  {\em \procspie} {\bf 9148},
   914810 (July 2014).

\bibitem{diolaiti10}
{Diolaiti}, E., {Conan}, J.-M., {Foppiani}, I., {Marchetti}, E., {Baruffolo},
  A., {Bellazzini}, M., {Bregoli}, G., {Butler}, C.~R., {Ciliegi}, P.,
  {Cosentino}, G., {Delabre}, B., {Lombini}, M., {Petit}, C., {Robert}, C.,
  {Rossettini}, P., {Schreiber}, L., {Tomelleri}, R., {Biliotti}, V.,
  {D'Odorico}, S., {Fusco}, T., {Hubin}, N., and {Meimon}, S., ``{Conceptual
  design and performance of the multiconjugate adaptive optics module for the
  European Extremely Large Telescope},'' in [{\em Society of Photo-Optical
  Instrumentation Engineers (SPIE) Conference
  Series}{\nolinebreak\hspace{0.1em}]},  {\em Society of Photo-Optical
  Instrumentation Engineers (SPIE) Conference Series} {\bf 7736} (July 2010).

\bibitem{diolaiti16}
{Diolaiti}, . t. M.~c., ``{MAORY: adaptive optics module for the E-ELT},'' in
  [{\em This SPIE conference}{\nolinebreak\hspace{0.1em}]},  {\em This SPIE
  conference} (2016).

\bibitem{davies10}
{Davies}, R., {Ageorges}, N., {Barl}, L., {Bedin}, L.~R., {Bender}, R.,
  {Bernardi}, P., {Chapron}, F., {Clenet}, Y., {Deep}, A., {Deul}, E., {Drost},
  M., {Eisenhauer}, F., {Falomo}, R., {Fiorentino}, G., {F{\"o}rster
  Schreiber}, N.~M., {Gendron}, E., {Genzel}, R., {Gratadour}, D., {Greggio},
  L., {Grupp}, F., {Held}, E., {Herbst}, T., {Hess}, H.-J., {Hubert}, Z.,
  {Jahnke}, K., {Kuijken}, K., {Lutz}, D., {Magrin}, D., {Muschielok}, B.,
  {Navarro}, R., {Noyola}, E., {Paumard}, T., {Piotto}, G., {Ragazzoni}, R.,
  {Renzini}, A., {Rousset}, G., {Rix}, H.-W., {Saglia}, R., {Tacconi}, L.,
  {Thiel}, M., {Tolstoy}, E., {Trippe}, S., {Tromp}, N., {Valentijn}, E.~A.,
  {Verdoes Kleijn}, G., and {Wegner}, M., ``{MICADO: the E-ELT adaptive optics
  imaging camera},'' in [{\em Society of Photo-Optical Instrumentation
  Engineers (SPIE) Conference Series}{\nolinebreak\hspace{0.1em}]},  {\em
  Society of Photo-Optical Instrumentation Engineers (SPIE) Conference Series}
  {\bf 7735} (July 2010).

\bibitem{marchetti08}
{Marchetti}, E., {Brast}, R., {Delabre}, B., {Donaldson}, R., {Fedrigo}, E.,
  {Frank}, C., {Hubin}, N., {Kolb}, J., {Lizon}, J.-L., {Marchesi}, M.,
  {Oberti}, S., {Reiss}, R., {Soenke}, C., {Tordo}, S., {Baruffolo}, A.,
  {Bagnara}, P., {Amorim}, A., and {Lima}, J., ``{MAD on sky results in star
  oriented mode},'' in [{\em Society of Photo-Optical Instrumentation Engineers
  (SPIE) Conference Series}{\nolinebreak\hspace{0.1em}]},  {\em Society of
  Photo-Optical Instrumentation Engineers (SPIE) Conference Series} {\bf 7015}
  (July 2008).

\bibitem{bouy08}
{Bouy}, H., {Kolb}, J., {Marchetti}, E., {Mart{\'{\i}}n}, E.~L., {Hu{\'e}lamo},
  N., and {Barrado Y Navascu{\'e}s}, D., ``{Multi-conjugate adaptive optics
  images of the Trapezium cluster},'' {\em \aap}~{\bf 477},  681--690 (Jan.
  2008).

\bibitem{gullieuszik08}
{Gullieuszik}, M., {Greggio}, L., {Held}, E.~V., {Moretti}, A., {Arcidiacono},
  C., {Bagnara}, P., {Baruffolo}, A., {Diolaiti}, E., {Falomo}, R., {Farinato},
  J., {Lombini}, M., {Ragazzoni}, R., {Brast}, R., {Donaldson}, R., {Kolb}, J.,
  {Marchetti}, E., and {Tordo}, S., ``{Resolving stellar populations outside
  the Local Group: MAD observations of UKS 2323-326},'' {\em \aap}~{\bf 483},
  L5--L8 (May 2008).

\bibitem{bouy09}
{Bouy}, H., {Hu{\'e}lamo}, N., {Mart{\'{\i}}n}, E.~L., {Marchis}, F., {Barrado
  Y Navascu{\'e}s}, D., {Kolb}, J., {Marchetti}, E., {Petr-Gotzens}, M.~G.,
  {Sterzik}, M., {Ivanov}, V.~D., {K{\"o}hler}, R., and {N{\"u}rnberger}, D.,
  ``{A deep look into the cores of young clusters. I. {$\sigma$}-Orionis},''
  {\em \aap}~{\bf 493},  931--946 (Jan. 2009).

\bibitem{ferraro09a}
{Ferraro}, F.~R., {Dalessandro}, E., {Mucciarelli}, A., {Beccari}, G., {Rich},
  R.~M., {Origlia}, L., {Lanzoni}, B., {Rood}, R.~T., {Valenti}, E.,
  {Bellazzini}, M., {Ransom}, S.~M., and {Cocozza}, G., ``{The cluster Terzan 5
  as a remnant of a primordial building block of the Galactic bulge},'' {\em
  \nat}~{\bf 462},  483--486 (Nov. 2009).

\bibitem{morettia09}
{Moretti}, A., {Piotto}, G., {Arcidiacono}, C., {Milone}, A.~P., {Ragazzoni},
  R., {Falomo}, R., {Farinato}, J., {Bedin}, L.~R., {Anderson}, J.,
  {Sarajedini}, A., {Baruffolo}, A., {Diolaiti}, E., {Lombini}, M., {Brast},
  R., {Donaldson}, R., {Kolb}, J., {Marchetti}, E., and {Tordo}, S., ``{MCAO
  near-IR photometry of the globular cluster NGC 6388: MAD observations in
  crowded fields},'' {\em \aap}~{\bf 493},  539--546 (Jan. 2009).

\bibitem{bono10a}
{Bono}, G., {Stetson}, P.~B., {VandenBerg}, D.~A., {Calamida}, A., {Dall'Ora},
  M., {Iannicola}, G., {Amico}, P., {Di Cecco}, A., {Marchetti}, E., {Monelli},
  M., {Sanna}, N., {Walker}, A.~R., {Zoccali}, M., {Buonanno}, R., {Caputo},
  F., {Corsi}, C.~E., {Degl'Innocenti}, S., {D'Odorico}, S., {Ferraro}, I.,
  {Gilmozzi}, R., {Melnick}, J., {Nonino}, M., {Ortolani}, S., {Piersimoni},
  A.~M., {Prada Moroni}, P.~G., {Pulone}, L., {Romaniello}, M., and {Storm},
  J., ``{On a New Near-Infrared Method to Estimate the Absolute Ages of Star
  Clusters: NGC 3201 as a First Test Case},'' {\em \apjl}~{\bf 708},  L74--L79
  (Jan. 2010).

\bibitem{campbell10}
{Campbell}, M.~A., {Evans}, C.~J., {Mackey}, A.~D., {Gieles}, M., {Alves}, J.,
  {Ascenso}, J., {Bastian}, N., and {Longmore}, A.~J., ``{VLT-MAD observations
  of the core of 30 Doradus},'' {\em \mnras}~{\bf 405},  421--435 (June 2010).

\bibitem{crowther10}
{Crowther}, P.~A., {Schnurr}, O., {Hirschi}, R., {Yusof}, N., {Parker}, R.~J.,
  {Goodwin}, S.~P., and {Kassim}, H.~A., ``{The R136 star cluster hosts several
  stars whose individual masses greatly exceed the accepted 150M$_{solar}$
  stellar mass limit},'' {\em \mnras}~{\bf 408},  731--751 (Oct. 2010).

\bibitem{sana10}
{Sana}, H., {Momany}, Y., {Gieles}, M., {Carraro}, G., {Beletsky}, Y.,
  {Ivanov}, V.~D., {de Silva}, G., and {James}, G., ``{A MAD view of Trumpler
  14},'' {\em \aap}~{\bf 515},  A26 (June 2010).

\bibitem{fiorentino11}
{Fiorentino}, G., {Tolstoy}, E., {Diolaiti}, E., {Valenti}, E., {Cignoni}, M.,
  and {Mackey}, A.~D., ``{MAD about the Large Magellanic Cloud. Preparing for
  the era of Extremely Large Telescopes},'' {\em \aap}~{\bf 535},  A63 (Nov.
  2011).

\bibitem{ortolani11}
{Ortolani}, S., {Barbuy}, B., {Momany}, Y., {Saviane}, I., {Bica}, E.,
  {Jilkova}, L., {Salerno}, G.~M., and {Jungwiert}, B., ``{A Fossil Bulge
  Globular Cluster Revealed by very Large Telescope Multi-conjugate Adaptive
  Optics},'' {\em \apj}~{\bf 737},  31 (Aug. 2011).

\bibitem{meyer11}
{Meyer}, E., {K{\"u}rster}, M., {Arcidiacono}, C., {Ragazzoni}, R., and {Rix},
  H.-W., ``{Astrometry with the MCAO instrument MAD. An analysis of
  single-epoch data obtained in the layer-oriented mode},'' {\em \aap}~{\bf
  532},  A16 (Aug. 2011).

\bibitem{rochau11}
{Rochau}, B., {Brandner}, W., {Stolte}, A., {Henning}, T., {Da Rio}, N.,
  {Gennaro}, M., {Hormuth}, F., {Marchetti}, E., and {Amico}, P., ``{A
  benchmark for multiconjugated adaptive optics: VLT-MAD observations of the
  young massive cluster Trumpler 14},'' {\em \mnras}~{\bf 418},  949--959 (Dec.
  2011).

\bibitem{mignani08}
{Mignani}, R.~P., {Falomo}, R., {Moretti}, A., {Treves}, A., {Turolla}, R.,
  {Sartore}, N., {Zane}, S., {Arcidiacono}, C., {Lombini}, M., {Farinato}, J.,
  {Baruffolo}, A., {Ragazzoni}, R., and {Marchetti}, E., ``{Near infrared
  VLT/MAD observations of the isolated neutron stars RX J0420.0-5022 and RX
  J1856.5-3754},'' {\em \aap}~{\bf 488},  267--270 (Sept. 2008).

\bibitem{falomo09}
{Falomo}, R., {Pian}, E., {Treves}, A., {Giovannini}, G., {Venturi}, T.,
  {Moretti}, A., {Arcidiacono}, C., {Farinato}, J., {Ragazzoni}, R.,
  {Diolaiti}, E., {Lombini}, M., {Tavecchio}, F., {Brast}, R., {Donaldson}, R.,
  {Kolb}, J., {Marchetti}, E., and {Tordo}, S., ``{The jet of the BL Lacertae
  object PKS 0521-365 in the near-IR: MAD adaptive optics observations},'' {\em
  \aap}~{\bf 501},  907--914 (July 2009).

\bibitem{melnick12}
{Melnick}, J., {Marchetti}, E., and {Amico}, P., ``{Science with ESO's
  Multi-conjugate Adaptive-optics Demonstrator - MAD},'' in [{\em Adaptive
  Optics Systems III}{\nolinebreak\hspace{0.1em}]},  {\em \procspie} {\bf
  8447},  84470M (July 2012).

\bibitem{stetson94}
{Stetson}, P.~B., ``{The center of the core-cusp globular cluster M15: CFHT and
  HST Observations, ALLFRAME reductions},'' {\em \pasp}~{\bf 106},  250--280
  (Mar. 1994).

\bibitem{rigaut12}
{Rigaut}, F., {Neichel}, B., {Boccas}, M., {d'Orgeville}, C., {Arriagada}, G.,
  {Fesquet}, V., {Diggs}, S.~J., {Marchant}, C., {Gausach}, G., {Rambold},
  W.~N., {Luhrs}, J., {Walker}, S., {Carrasco-Damele}, E.~R., {Edwards}, M.~L.,
  {Pessev}, P., {Galvez}, R.~L., {Vucina}, T.~B., {Araya}, C., {Gutierrez}, A.,
  {Ebbers}, A.~W., {Serio}, A., {Moreno}, C., {Urrutia}, C., {Rogers}, R.,
  {Rojas}, R., {Trujillo}, C., {Miller}, B., {Simons}, D.~A., {Lopez}, A.,
  {Montes}, V., {Diaz}, H., {Daruich}, F., {Colazo}, F., {Bec}, M., {Trancho},
  G., {Sheehan}, M., {McGregor}, P., {Young}, P.~J., {Doolan}, M.~C., {van
  Harmelen}, J., {Ellerbroek}, B.~L., {Gratadour}, D., and {Garcia-Rissmann},
  A., ``{GeMS: first on-sky results},'' in [{\em Adaptive Optics Systems
  III}{\nolinebreak\hspace{0.1em}]},  {\em \procspie} {\bf 8447},  84470I (July
  2012).

\bibitem{rigaut14}
{Rigaut}, F., {Neichel}, B., {Boccas}, M., {d'Orgeville}, C., {Vidal}, F., {van
  Dam}, M.~A., {Arriagada}, G., {Fesquet}, V., {Galvez}, R.~L., {Gausachs}, G.,
  {Cavedoni}, C., {Ebbers}, A.~W., {Karewicz}, S., {James}, E., {L{\"u}hrs},
  J., {Montes}, V., {Perez}, G., {Rambold}, W.~N., {Rojas}, R., {Walker}, S.,
  {Bec}, M., {Trancho}, G., {Sheehan}, M., {Irarrazaval}, B., {Boyer}, C.,
  {Ellerbroek}, B.~L., {Flicker}, R., {Gratadour}, D., {Garcia-Rissmann}, A.,
  and {Daruich}, F., ``{Gemini multiconjugate adaptive optics system review -
  I. Design, trade-offs and integration},'' {\em \mnras}~{\bf 437},  2361--2375
  (Jan. 2014).

\bibitem{neichel14a}
{Neichel}, B., {Vidal}, F., {Rigaut}, F., {Rodrigo Carrasco}, E., {Arriagada},
  G., {Serio}, A., {Pessev}, P., {Winge}, C., {van Dam}, M., {Garrel}, V.,
  {Araujo}, C., {Boccas}, M., {Fesquet}, V., {Galvez}, R., {Gausachs}, G.,
  {Luhrs}, J., {Montes}, V., {Moreno}, C., {Rambold}, W., {Trujillo}, C.,
  {Urrutia}, C., and {Vucina}, T., ``{Gems first science results},'' {\em ArXiv
  e-prints}  (Jan. 2014).

\bibitem{davidge14}
{Davidge}, T.~J., ``{GeMS in the Outer Galaxy: Near-infrared Imaging of Three
  Young Clusters at Large Galactic Radii},'' {\em \apj}~{\bf 781},  95 (Feb.
  2014).

\bibitem{saracino15}
{Saracino}, S., {Dalessandro}, E., {Ferraro}, F.~R., {Lanzoni}, B., {Geisler},
  D., {Mauro}, F., {Villanova}, S., {Moni Bidin}, C., {Miocchi}, P., and
  {Massari}, D., ``{GEMINI/GeMS Observations Unveil the Structure of the
  Heavily Obscured Globular Cluster Liller 1.},'' {\em \apj}~{\bf 806},  152
  (June 2015).

\bibitem{manchado15}
{Manchado}, A., {Stanghellini}, L., {Villaver}, E., {Garc{\'{\i}}a-Segura}, G.,
  {Shaw}, R.~A., and {Garc{\'{\i}}a-Hern{\'a}ndez}, D.~A., ``{High-resolution
  Imaging of NGC 2346 with GSAOI/GeMS: Disentangling the Planetary Nebula
  Molecular Structure to Understand Its Origin and Evolution},'' {\em
  \apj}~{\bf 808},  115 (Aug. 2015).

\bibitem{turri15}
{Turri}, P., {McConnachie}, A.~W., {Stetson}, P.~B., {Fiorentino}, G.,
  {Andersen}, D.~R., {V{\'e}ran}, J.-P., and {Bono}, G., ``{Toward Precision
  Photometry for the ELT Era: The Double Subgiant Branch of NGC 1851 Observed
  with the Gemini/GeMS MCAO System},'' {\em \apjl}~{\bf 811},  L15 (Oct. 2015).

\bibitem{massari16a}
{Massari}, D., {Fiorentino}, G., {McConnachie}, A., {Bono}, G., {Dall'Ora}, M.,
  {Ferraro}, I., {Iannicola}, G., {Stetson}, P.~B., {Turri}, P., and {Tolstoy},
  E., ``{GeMS MCAO observations of the Galactic globular cluster NGC 2808: the
  absolute age},'' {\em \aap}~{\bf 586},  A51 (Feb. 2016).

\bibitem{santos16}
{Santos}, Jr., J.~F.~C., {Roman-Lopes}, A., {Carrasco}, E.~R., {Maia},
  F.~F.~S., and {Neichel}, B., ``{GeMs/GSAOI observations of La Serena 94: an
  old and far open cluster inside the solar circle},'' {\em \mnras}~{\bf 456},
  2126--2139 (Feb. 2016).

\bibitem{opitz16}
{Opitz}, D., {Tinney}, C.~G., {Faherty}, J.~K., {Sweet}, S., {Gelino}, C.~R.,
  and {Kirkpatrick}, J.~D., ``{Searching for Binary Y Dwarfs with the Gemini
  Multi-conjugate Adaptive Optics System (GeMS)},'' {\em \apj}~{\bf 819},  17
  (Mar. 2016).

\bibitem{bernard16}
{Bernard}, A., {Neichel}, B., {Samal}, M.~R., {Zavagno}, A., {Andersen}, M.,
  {Evans}, C.~J., {Plana}, H., and {Fusco}, T., ``{Deep GeMS/GSAOI
  near-infrared observations of N159W in the Large Magellanic Cloud},'' {\em
  ArXiv e-prints}  (May 2016).

\bibitem{esposito12}
{Esposito}, S., {Riccardi}, A., {Pinna}, E., {Puglisi}, A.~T.,
  {Quir{\'o}s-Pacheco}, F., {Arcidiacono}, C., {Xompero}, M., {Briguglio}, R.,
  {Busoni}, L., {Fini}, L., {Argomedo}, J., {Gherardi}, A., {Agapito}, G.,
  {Brusa}, G., {Miller}, D.~L., {Guerra Ramon}, J.~C., {Boutsia}, K., and
  {Stefanini}, P., ``{Natural guide star adaptive optics systems at LBT: FLAO
  commissioning and science operations status},'' in [{\em Adaptive Optics
  Systems III}{\nolinebreak\hspace{0.1em}]},  {\em \procspie} {\bf 8447},
  84470U (July 2012).

\bibitem{dicecco15b}
{Di Cecco}, A., {Bono}, G., {Prada Moroni}, P.~G., {Tognelli}, E., {Allard},
  F., {Stetson}, P.~B., {Buonanno}, R., {Ferraro}, I., {Iannicola}, G.,
  {Monelli}, M., {Nonino}, M., and {Pulone}, L., ``{On the Absolute Age of the
  Metal-rich Globular M71 (NGC 6838). I. Optical Photometry},'' {\em \aj}~{\bf
  150},  51 (Aug. 2015).

\bibitem{monelli15}
{Monelli}, M., {Testa}, V., {Bono}, G., {Ferraro}, I., {Iannicola}, G.,
  {Fiorentino}, G., {Arcidiacono}, C., {Massari}, D., {Boutsia}, K.,
  {Briguglio}, R., {Busoni}, L., {Carini}, R., {Close}, L., {Cresci}, G.,
  {Esposito}, S., {Fini}, L., {Fumana}, M., {Guerra}, J.~C., {Hill}, J.,
  {Kulesa}, C., {Mannucci}, F., {McCarthy}, D., {Pinna}, E., {Puglisi}, A.,
  {Quiros-Pacheco}, F., {Ragazzoni}, R., {Riccardi}, A., {Skemer}, A., and
  {Xompero}, M., ``{The Absolute Age of the Globular Cluster M15 Using
  Near-infrared Adaptive Optics Images from PISCES/LBT.},'' {\em \apj}~{\bf
  812},  25 (Oct. 2015).

\bibitem{massari16c}
{Massari}, D., {Fiorentino}, G., {Tolstoy}, E., {McConnachie}, A., {Stuik}, R.,
  {Schreiber}, L., {Andersen}, D., {Cl{\'e}net}, Y., {Davies}, R., {Gratadour},
  D., {Kuijken}, K., {Navarro}, R., {Pott}, J.-U., {Rodeghiero}, G., {Turri},
  P., and {Verdoes Kleijn}, G., ``{High-precision astrometry towards ELTs},''
  {\em ArXiv e-prints}  (July 2016).

\bibitem{neichel14b}
{Neichel}, B., {Lu}, J.~R., {Rigaut}, F., {Ammons}, S.~M., {Carrasco}, E.~R.,
  and {Lassalle}, E., ``{Astrometric performance of the Gemini multiconjugate
  adaptive optics system in crowded fields},'' {\em \mnras}~{\bf 445},
  500--514 (Nov. 2014).

\bibitem{eisenstein05}
{Eisenstein}, D.~J., {Zehavi}, I., {Hogg}, D.~W., {Scoccimarro}, R., {Blanton},
  M.~R., {Nichol}, R.~C., {Scranton}, R., {Seo}, H.-J., {Tegmark}, M., {Zheng},
  Z., {Anderson}, S.~F., {Annis}, J., {Bahcall}, N., {Brinkmann}, J., {Burles},
  S., {Castander}, F.~J., {Connolly}, A., {Csabai}, I., {Doi}, M., {Fukugita},
  M., {Frieman}, J.~A., {Glazebrook}, K., {Gunn}, J.~E., {Hendry}, J.~S.,
  {Hennessy}, G., {Ivezi{\'c}}, Z., {Kent}, S., {Knapp}, G.~R., {Lin}, H.,
  {Loh}, Y.-S., {Lupton}, R.~H., {Margon}, B., {McKay}, T.~A., {Meiksin}, A.,
  {Munn}, J.~A., {Pope}, A., {Richmond}, M.~W., {Schlegel}, D., {Schneider},
  D.~P., {Shimasaku}, K., {Stoughton}, C., {Strauss}, M.~A., {SubbaRao}, M.,
  {Szalay}, A.~S., {Szapudi}, I., {Tucker}, D.~L., {Yanny}, B., and {York},
  D.~G., ``{Detection of the Baryon Acoustic Peak in the Large-Scale
  Correlation Function of SDSS Luminous Red Galaxies},'' {\em \apj}~{\bf 633},
  560--574 (Nov. 2005).

\bibitem{riess16}
{Riess}, A.~G., {Macri}, L.~M., {Hoffmann}, S.~L., {Scolnic}, D., {Casertano},
  S., {Filippenko}, A.~V., {Tucker}, B.~E., {Reid}, M.~J., {Jones}, D.~O.,
  {Silverman}, J.~M., {Chornock}, R., {Challis}, P., {Yuan}, W., {Brown},
  P.~J., and {Foley}, R.~J., ``{A 2.4\% Determination of the Local Value of the
  Hubble Constant},'' {\em ArXiv e-prints}  (Apr. 2016).

\bibitem{riess11a}
{Riess}, A.~G., {Macri}, L., {Casertano}, S., {Lampeitl}, H., {Ferguson},
  H.~C., {Filippenko}, A.~V., {Jha}, S.~W., {Li}, W., and {Chornock}, R., ``{A
  3\% Solution: Determination of the Hubble Constant with the Hubble Space
  Telescope and Wide Field Camera 3},'' {\em \apj}~{\bf 730},  119 (Apr. 2011).

\bibitem{suyu13}
{Suyu}, S.~H., {Auger}, M.~W., {Hilbert}, S., {Marshall}, P.~J., {Tewes}, M.,
  {Treu}, T., {Fassnacht}, C.~D., {Koopmans}, L.~V.~E., {Sluse}, D.,
  {Blandford}, R.~D., {Courbin}, F., and {Meylan}, G., ``{Two Accurate
  Time-delay Distances from Strong Lensing: Implications for Cosmology},'' {\em
  \apj}~{\bf 766},  70 (Apr. 2013).

\bibitem{bennett14}
{Bennett}, C.~L., {Larson}, D., {Weiland}, J.~L., and {Hinshaw}, G., ``{The 1\%
  Concordance Hubble Constant},'' {\em \apj}~{\bf 794},  135 (Oct. 2014).

\bibitem{hinshaw13}
{Hinshaw}, G., {Larson}, D., {Komatsu}, E., {Spergel}, D.~N., {Bennett}, C.~L.,
  {Dunkley}, J., {Nolta}, M.~R., {Halpern}, M., {Hill}, R.~S., {Odegard}, N.,
  {Page}, L., {Smith}, K.~M., {Weiland}, J.~L., {Gold}, B., {Jarosik}, N.,
  {Kogut}, A., {Limon}, M., {Meyer}, S.~S., {Tucker}, G.~S., {Wollack}, E., and
  {Wright}, E.~L., ``{Nine-year Wilkinson Microwave Anisotropy Probe (WMAP)
  Observations: Cosmological Parameter Results},'' {\em \apjs}~{\bf 208},  19
  (Oct. 2013).

\bibitem{planck15}
{Planck Collaboration}, {Ade}, P.~A.~R., {Aghanim}, N., {Arnaud}, M.,
  {Ashdown}, M., {Aumont}, J., {Baccigalupi}, C., {Banday}, A.~J., {Barreiro},
  R.~B., {Battaner}, E., {Benabed}, K., {Benoit-L{\'e}vy}, A., {Bernard},
  J.-P., {Bersanelli}, M., {Bielewicz}, P., {Bond}, J.~R., {Borrill}, J.,
  {Bouchet}, F.~R., {Burigana}, C., {Butler}, R.~C., {Calabrese}, E.,
  {Chamballu}, A., {Chiang}, H.~C., {Christensen}, P.~R., {Clements}, D.~L.,
  {Colombo}, L.~P.~L., {Couchot}, F., {Curto}, A., {Cuttaia}, F., {Danese}, L.,
  {Davies}, R.~D., {Davis}, R.~J., {de Bernardis}, P., {de Rosa}, A., {de
  Zotti}, G., {Delabrouille}, J., {Diego}, J.~M., {Dole}, H., {Dor{\'e}}, O.,
  {Dupac}, X., {En{\ss}lin}, T.~A., {Eriksen}, H.~K., {Fabre}, O., {Finelli},
  F., {Forni}, O., {Frailis}, M., {Franceschi}, E., {Galeotta}, S., {Galli},
  S., {Ganga}, K., {Giard}, M., {Gonz{\'a}lez-Nuevo}, J., {G{\'o}rski}, K.~M.,
  {Gregorio}, A., {Gruppuso}, A., {Hansen}, F.~K., {Hanson}, D., {Harrison},
  D.~L., {Henrot-Versill{\'e}}, S., {Hern{\'a}ndez-Monteagudo}, C., {Herranz},
  D., {Hildebrandt}, S.~R., {Hivon}, E., {Hobson}, M., {Holmes}, W.~A.,
  {Hornstrup}, A., {Hovest}, W., {Huffenberger}, K.~M., {Jaffe}, A.~H.,
  {Jones}, W.~C., {Keih{\"a}nen}, E., {Keskitalo}, R., {Kneissl}, R., {Knoche},
  J., {Kunz}, M., {Kurki-Suonio}, H., {Lamarre}, J.-M., {Lasenby}, A.,
  {Lawrence}, C.~R., {Leonardi}, R., {Lesgourgues}, J., {Liguori}, M., {Lilje},
  P.~B., {Linden-V{\o}rnle}, M., {L{\'o}pez-Caniego}, M., {Lubin}, P.~M.,
  {Mac{\'{\i}}as-P{\'e}rez}, J.~F., {Mandolesi}, N., {Maris}, M., {Martin},
  P.~G., {Mart{\'{\i}}nez-Gonz{\'a}lez}, E., {Masi}, S., {Matarrese}, S.,
  {Mazzotta}, P., {Meinhold}, P.~R., {Melchiorri}, A., {Mendes}, L.,
  {Menegoni}, E., {Mennella}, A., {Migliaccio}, M., {Miville-Desch{\^e}nes},
  M.-A., {Moneti}, A., {Montier}, L., {Morgante}, G., {Moss}, A., {Munshi}, D.,
  {Murphy}, J.~A., {Naselsky}, P., {Nati}, F., {Natoli}, P.,
  {N{\o}rgaard-Nielsen}, H.~U., {Noviello}, F., {Novikov}, D., {Novikov}, I.,
  {Oxborrow}, C.~A., {Pagano}, L., {Pajot}, F., {Paoletti}, D., {Pasian}, F.,
  {Patanchon}, G., {Perdereau}, O., {Perotto}, L., {Perrotta}, F.,
  {Piacentini}, F., {Piat}, M., {Pierpaoli}, E., {Pietrobon}, D.,
  {Plaszczynski}, S., {Pointecouteau}, E., {Polenta}, G., {Ponthieu}, N.,
  {Popa}, L., {Pratt}, G.~W., {Prunet}, S., {Rachen}, J.~P., {Rebolo}, R.,
  {Reinecke}, M., {Remazeilles}, M., {Renault}, C., {Ricciardi}, S.,
  {Ristorcelli}, I., {Rocha}, G., {Roudier}, G., {Rusholme}, B., {Sandri}, M.,
  {Savini}, G., {Scott}, D., {Spencer}, L.~D., {Stolyarov}, V., {Sudiwala}, R.,
  {Sutton}, D., {Suur-Uski}, A.-S., {Sygnet}, J.-F., {Tauber}, J.~A.,
  {Tavagnacco}, D., {Terenzi}, L., {Toffolatti}, L., {Tomasi}, M., {Tristram},
  M., {Tucci}, M., {Uzan}, J.-P., {Valenziano}, L., {Valiviita}, J., {Van
  Tent}, B., {Vielva}, P., {Villa}, F., {Wade}, L.~A., {Yvon}, D., {Zacchei},
  A., and {Zonca}, A., ``{Planck intermediate results. XXIV. Constraints on
  variations in fundamental constants},'' {\em \aap}~{\bf 580},  A22 (Aug.
  2015).

\bibitem{cuesta16}
{Cuesta}, A.~J., {Vargas-Maga{\~n}a}, M., {Beutler}, F., {Bolton}, A.~S.,
  {Brownstein}, J.~R., {Eisenstein}, D.~J., {Gil-Mar{\'{\i}}n}, H., {Ho}, S.,
  {McBride}, C.~K., {Maraston}, C., {Padmanabhan}, N., {Percival}, W.~J.,
  {Reid}, B.~A., {Ross}, A.~J., {Ross}, N.~P., {S{\'a}nchez}, A.~G.,
  {Schlegel}, D.~J., {Schneider}, D.~P., {Thomas}, D., {Tinker}, J., {Tojeiro},
  R., {Verde}, L., and {White}, M., ``{The clustering of galaxies in the
  SDSS-III Baryon Oscillation Spectroscopic Survey: baryon acoustic
  oscillations in the correlation function of LOWZ and CMASS galaxies in Data
  Release 12},'' {\em \mnras}~{\bf 457},  1770--1785 (Apr. 2016).

\bibitem{freedman10}
{Freedman}, W.~L. and {Madore}, B.~F., ``{The Hubble Constant},'' {\em
  \araa}~{\bf 48},  673--710 (Sept. 2010).

\bibitem{leaman13}
{Leaman}, R., {VandenBerg}, D.~A., and {Mendel}, J.~T., ``{The bifurcated
  age-metallicity relation of Milky Way globular clusters and its implications
  for the accretion history of the galaxy},'' {\em \mnras}~{\bf 436},  122--135
  (Nov. 2013).

\bibitem{marinfranch09}
{Mar{\'{\i}}n-Franch}, A., {Aparicio}, A., {Piotto}, G., {Rosenberg}, A.,
  {Chaboyer}, B., {Sarajedini}, A., {Siegel}, M., {Anderson}, J., {Bedin},
  L.~R., {Dotter}, A., {Hempel}, M., {King}, I., {Majewski}, S., {Milone},
  A.~P., {Paust}, N., and {Reid}, I.~N., ``{The ACS Survey of Galactic Globular
  Clusters. VII. Relative Ages},'' {\em \apj}~{\bf 694},  1498--1516 (Apr.
  2009).

\bibitem{vandenberg13}
{VandenBerg}, D.~A., {Brogaard}, K., {Leaman}, R., and {Casagrande}, L., ``{The
  Ages of 55 Globular Clusters as Determined Using an Improved
  $\backslash$Delta V\^{}HB\_TO Method along with Color-Magnitude Diagram
  Constraints, and Their Implications for Broader Issues},'' {\em \apj}~{\bf
  775},  134 (Oct. 2013).

\bibitem{osborn71}
{Osborn}, W.~H., {\em {Positions of Globular Cluster Stars in the Physical H-R
  Diagram.}}, PhD thesis, Yale University. (1971).

\bibitem{gratton12}
{Gratton}, R.~G., {Carretta}, E., and {Bragaglia}, A., ``{Multiple populations
  in globular clusters. Lessons learned from the Milky Way globular
  clusters},'' {\em \aapr}~{\bf 20},  50 (Feb. 2012).

\bibitem{bellini13}
{Bellini}, A., {Piotto}, G., {Milone}, A.~P., {King}, I.~R., {Renzini}, A.,
  {Cassisi}, S., {Anderson}, J., {Bedin}, L.~R., {Nardiello}, D.,
  {Pietrinferni}, A., and {Sarajedini}, A., ``{The Intriguing Stellar
  Populations in the Globular Clusters NGC 6388 and NGC 6441},'' {\em
  \apj}~{\bf 765},  32 (Mar. 2013).

\bibitem{milone14}
{Milone}, A.~P., {Marino}, A.~F., {Bedin}, L.~R., {Piotto}, G., {Cassisi}, S.,
  {Dieball}, A., {Anderson}, J., {Jerjen}, H., {Asplund}, M., {Bellini}, A.,
  {Brogaard}, K., {Dotter}, A., {Giersz}, M., {Heggie}, D.~C., {Knigge}, C.,
  {Rich}, R.~M., {van den Berg}, M., and {Buonanno}, R., ``{The M 4 Core
  Project with HST - II. Multiple stellar populations at the bottom of the main
  sequence},'' {\em \mnras}~{\bf 439},  1588--1595 (Apr. 2014).

\bibitem{ventura01}
{Ventura}, P., {D'Antona}, F., {Mazzitelli}, I., and {Gratton}, R.,
  ``{Predictions for Self-Pollution in Globular Cluster Stars},'' {\em
  \apjl}~{\bf 550},  L65--L69 (Mar. 2001).

\bibitem{pulone98}
{Pulone}, L., {De Marchi}, G., {Paresce}, F., and {Allard}, F., ``{The Lower
  Main Sequence of {$\omega$} Centauri from Deep Hubble Space Telescope NICMOS
  Near-Infrared Observations},'' {\em \apjl}~{\bf 492},  L41--L44 (Jan. 1998).

\bibitem{pulone99}
{Pulone}, L., {de Marchi}, G., and {Paresce}, F., ``{The mass function of M 4
  from near IR and optical HST observations},'' {\em \aap}~{\bf 342},  440--452
  (Feb. 1999).

\bibitem{lagioia14}
{Lagioia}, E.~P., {Milone}, A.~P., {Stetson}, P.~B., {Bono}, G., {Prada
  Moroni}, P.~G., {Dall'Ora}, M., {Aparicio}, A., {Buonanno}, R., {Calamida},
  A., {Ferraro}, I., {Gilmozzi}, R., {Iannicola}, G., {Matsunaga}, N.,
  {Monelli}, M., and {Walker}, A., ``{On the Kinematic Separation of Field and
  Cluster Stars across the Bulge Globular NGC 6528},'' {\em \apj}~{\bf 782},
  50 (Feb. 2014).

\bibitem{milone12}
{Milone}, A.~P., {Marino}, A.~F., {Cassisi}, S., {Piotto}, G., {Bedin}, L.~R.,
  {Anderson}, J., {Allard}, F., {Aparicio}, A., {Bellini}, A., {Buonanno}, R.,
  {Monelli}, M., and {Pietrinferni}, A., ``{The infrared eye of the Wide-Field
  Camera 3 on the Hubble Space Telescope reveals multiple main sequences of
  very low-mass stars in NGC 2808},'' {\em ArXiv e-prints}  (June 2012).

\bibitem{turri14a}
{Turri}, P., {McConnachie}, A.~W., {Stetson}, P.~B., {Fiorentino}, G.,
  {Andersen}, D.~R., {Bono}, G., and {V{\'e}ran}, J.-P., ``{Photometric
  performance of LGS MCAO with science-based metrics: first results from
  Gemini/GeMS observations of Galactic globular clusters},'' in [{\em Adaptive
  Optics Systems IV}{\nolinebreak\hspace{0.1em}]},  {\em \procspie} {\bf 9148},
   91483V (Aug. 2014).

\bibitem{correnti16}
{Correnti}, M., {Gennaro}, M., {Kalirai}, J.~S., {Brown}, T.~M., and
  {Calamida}, A., ``{Constraining Globular Cluster Age Uncertainties using the
  IR Color--Magnitude Diagram},'' {\em \apj}~{\bf 823},  18 (May 2016).

\bibitem{sarajedini09a}
{Sarajedini}, A., {Dotter}, A., and {Kirkpatrick}, A., ``{Deep 2MASS Photometry
  of M67 and Calibration of the Main-Sequence J - K$_{S}$ Color Difference as
  an Age Indicator},'' {\em \apj}~{\bf 698},  1872--1878 (June 2009).

\bibitem{zoccali00a}
{Zoccali}, M., {Cassisi}, S., {Frogel}, J.~A., {Gould}, A., {Ortolani}, S.,
  {Renzini}, A., {Rich}, R.~M., and {Stephens}, A.~W., ``{The Initial Mass
  Function of the Galactic Bulge down to \~{}0.15 M$_{solar}$},'' {\em
  \apj}~{\bf 530},  418--428 (Feb. 2000).

\bibitem{saumon94}
{Saumon}, D., {Bergeron}, P., {Lunine}, J.~I., {Hubbard}, W.~B., and {Burrows},
  A., ``{Cool zero-metallicity stellar atmospheres},'' {\em \apj}~{\bf 424},
  333--344 (Mar. 1994).

\bibitem{borysow97}
{Borysow}, A., {Jorgensen}, U.~G., and {Zheng}, C., ``{Model atmospheres of
  cool, low-metallicity stars: the importance of collision-induced
  absorption.},'' {\em \aap}~{\bf 324},  185--195 (Aug. 1997).

\bibitem{allard95}
{Allard}, F. and {Hauschildt}, P.~H., ``{Model atmospheres for M (sub)dwarf
  stars. 1: The base model grid},'' {\em \apj}~{\bf 445},  433--450 (May 1995).

\bibitem{salaris16}
{Salaris}, M., {Cassisi}, S., and {Pietrinferni}, A., ``{On the red giant
  branch mass loss in 47 Tucanae: Constraints from the horizontal branch
  morphology},'' {\em \aap}~{\bf 590},  A64 (May 2016).

\bibitem{schreiber11}
{Schreiber}, L., {Diolaiti}, E., {Bellazzini}, M., {Ciliegi}, P., {Foppiani},
  I., {Greggio}, L., {Lanzoni}, B., and {Lombini}, M., ``{Handling a highly
  structured and spatially variable Point Spread Function in AO images},'' in
  [{\em Second International Conference on Adaptive Optics for Extremely Large
  Telescopes. Online at <A
  href=''http://ao4elt2.lesia.obspm.fr''>http://ao4elt2.lesia.obspm.fr</A>,
  id.P57}{\nolinebreak\hspace{0.1em}]},   P57 (Sept. 2011).

\bibitem{schreiber13}
{Schreiber}, L., {La Camera}, A., {Prato}, M., {Diolaiti}, E., {constraint on
  the PSFS which is an upper bound derived from the Strehl ratio (SR)}, h. s.
  i. g. w. s. P. a. f. t. P.-A. s. o. t. E.-E. M. s.~M., and different~crowding
  conditions., ``{Point Spread Function extraction in crowded fields using
  blind deconvolution},'' in [{\em Proceedings of the Third AO4ELT
  Conference}{\nolinebreak\hspace{0.1em}]},  {Esposito}, S. and {Fini}, L.,
  eds. (Dec. 2013).

\bibitem{fiorentino14b}
{Fiorentino}, G., {Ferraro}, I., {Iannicola}, G., {Bono}, G., {Monelli}, M.,
  {Testa}, V., {Arcidiacono}, C., {Faccini}, M., {Gilmozzi}, R., {Xompero}, M.,
  and {Briguglio}, R., ``{On the use of asymmetric PSF on NIR images of crowded
  stellar fields},'' in [{\em Society of Photo-Optical Instrumentation
  Engineers (SPIE) Conference Series}{\nolinebreak\hspace{0.1em}]},  {\em
  Society of Photo-Optical Instrumentation Engineers (SPIE) Conference Series}
  {\bf 9148},  3 (Aug. 2014).

\bibitem{milone08}
{Milone}, A.~P., {Bedin}, L.~R., {Piotto}, G., {Anderson}, J., {King}, I.~R.,
  {Sarajedini}, A., {Dotter}, A., {Chaboyer}, B., {Mar{\'{\i}}n-Franch}, A.,
  {Majewski}, S., {Aparicio}, A., {Hempel}, M., {Paust}, N.~E.~Q., {Reid},
  I.~N., {Rosenberg}, A., and {Siegel}, M., ``{The ACS Survey of Galactic
  Globular Clusters. III. The Double Subgiant Branch of NGC 1851},'' {\em
  \apj}~{\bf 673},  241--250 (Jan. 2008).

\bibitem{cassisi08}
{Cassisi}, S., {Salaris}, M., {Pietrinferni}, A., {Piotto}, G., {Milone},
  A.~P., {Bedin}, L.~R., and {Anderson}, J., ``{The Double Subgiant Branch of
  NGC 1851: The Role of the CNO Abundance},'' {\em \apjl}~{\bf 672},  L115
  (Jan. 2008).

\bibitem{turri16}
{Turri}, P., {McConnachie}, A.~W., {Stetson}, P.~B., {Andersen}, D.~R.,
  {V{\'e}ran}, J., {Fiorentino}, G., and {Massari}, D., ``{Photometric
  techniques, performance and PSF characterization of GeMS},'' in [{\em This
  SPIE conference}{\nolinebreak\hspace{0.1em}]},  {\em This SPIE conference}
  (2016).

\bibitem{milone15}
{Milone}, A.~P., {Marino}, A.~F., {Piotto}, G., {Renzini}, A., {Bedin}, L.~R.,
  {Anderson}, J., {Cassisi}, S., {D'Antona}, F., {Bellini}, A., {Jerjen}, H.,
  {Pietrinferni}, A., and {Ventura}, P., ``{The Hubble Space Telescope UV
  Legacy Survey of Galactic Globular Clusters. III. A Quintuple Stellar
  Population in NGC 2808},'' {\em \apj}~{\bf 808},  51 (July 2015).

\bibitem{carretta15}
{Carretta}, E., ``{Five Groups of Red Giants with Distinct Chemical Composition
  in the Globular Cluster NGC 2808},'' {\em \apj}~{\bf 810},  148 (Sept. 2015).

\bibitem{gallart96b}
{Gallart}, C., {Aparicio}, A., {Bertelli}, G., and {Chiosi}, C., ``{The Local
  Group Dwarf Irregular Galaxy NGC 6822.II.The Old and Intermediate -Age Star
  Formation History},'' {\em \aj}~{\bf 112},  1950--+ (Nov. 1996).

\bibitem{deep11}
{Deep}, A., {Fiorentino}, G., {Tolstoy}, E., {Diolaiti}, E., {Bellazzini}, M.,
  {Ciliegi}, P., {Davies}, R.~I., and {Conan}, J.-M., ``{An E-ELT case study:
  colour-magnitude diagrams of an old galaxy in the Virgo cluster},'' {\em
  \aap}~{\bf 531},  A151 (July 2011).

\bibitem{diolaiti00}
{Diolaiti}, E., {Bendinelli}, O., {Bonaccini}, D., {Close}, L., {Currie}, D.,
  and {Parmeggiani}, G., ``{Analysis of isoplanatic high resolution stellar
  fields by the StarFinder code},'' {\em \aaps}~{\bf 147},  335--346 (Dec.
  2000).

\bibitem{lacamera15}
{La Camera}, A., {Schreiber}, L., {Diolaiti}, E., {Boccacci}, P., {Bertero},
  M., {Bellazzini}, M., and {Ciliegi}, P., ``{A method for space-variant
  deblurring with application to adaptive optics imaging in astronomy},'' {\em
  \aap}~{\bf 579},  A1 (July 2015).

\bibitem{buonanno89}
{Buonanno}, R. and {Iannicola}, G., ``{Stellar photometry with big pixels},''
  {\em \pasp}~{\bf 101},  294--301 (Mar. 1989).

\bibitem{jolissant14}
{Jolissaint}, L., {Ragland}, S., {Wizinowich}, P., and {Bouxin}, A., ``{Laser
  guide star adaptive optics point spread function reconstruction project at W.
  M. Keck Observatory: preliminary on-sky results},'' in [{\em Adaptive Optics
  Systems IV}{\nolinebreak\hspace{0.1em}]},  {\em \procspie} {\bf 9148},
  91484S (July 2014).

\end{thebibliography}
\bibliographystyle{spiebib} 

\end{document}